\documentclass[review]{elsarticle}
\usepackage[hyphens]{url}
\usepackage{hyperref}
\RequirePackage{amsthm,amssymb,amsmath}
\RequirePackage{subcaption}
\usepackage[usenames,dvipsnames,svgnames,table]{xcolor}

\usepackage{booktabs} % professional-quality tables
\RequirePackage{algorithm}
\RequirePackage{algorithmic}
\usepackage{setspace}
\usepackage{multirow}
\usepackage{textgreek} 
\usepackage{amsfonts}  % blackboard math symbols
\usepackage{bm}
\usepackage{nicefrac}  % compact symbols for 1/2, etc.
\usepackage{microtype}  % microtypography
\usepackage[algo2e, ruled, linesnumbered, resetcount]{algorithm2e}

\def\HiLi{\leavevmode\rlap{\hbox to \hsize{\color{yellow!50}\leaders\hrule height .8\baselineskip depth .5ex\hfill}}}

\makeatletter
\newcommand{\removelatexerror}{\let\@latex@error\@gobble}
\makeatother

\newcommand{\iust}{\textit{IUST-DeepFuzz}}
\newcommand{\iustpc}{\textit{IUST-PDFCorpus}}
\newcommand{\mdnf}{\textit{MetadataNeuralFuzz}}
\newcommand{\dnf}{\textit{DataNeuralFuzz}}

\journal{arXiv}

\begin{document}

\begin{frontmatter}

\title{Format-aware Learn\&Fuzz: Deep Test Data Generation for Efficient Fuzzing}%\tnoteref{mytitlenote}}

\author[mymainaddress]{Morteza Zakeri Nasrabadi}%\fnref{myfootnote}}
\ead{morteza\_zakeri@comp.iust.ac.ir}
%\address{Iran University of Science and Technology, Tehran, Iran.}

\author[mymainaddress]{Saeed Parsa\corref{mycorrespondingauthor}}
\cortext[mycorrespondingauthor]{Corresponding author}
%\ead[url]{www.parsa.iust.ac.ir}
\ead{parsa@iust.ac.ir}

\author[mymainaddress]{Akram Kalaee}
\ead{kalaee@comp.iust.ac.ir}

\address[mymainaddress]{Iran University of Science and Technology, Tehran, Iran.}
%\address[mysecondaryaddress]{Iran University of Science and Technology, Tehran, Iran.}

\begin{abstract}
Appropriate test data is a crucial factor to reach success in dynamic software testing, e.g., fuzzing. Most of the real-world applications, however, accept complex structure inputs containing data surrounded by meta-data which is processed in several stages comprising of the parsing and rendering (execution). It makes the automatically generating efficient test data, to be non-trivial and laborious activity. The success of deep learning to cope in solving complex tasks especially in generative tasks has motivated us to exploit it in the context of complex test data generation. To do so, a neural language model (NLM) based on deep recurrent neural networks (RNNs) is used to learn the structure of complex input. Our approach generates new test data while distinguishes between data and meta-data that makes it possible to target both the parsing and rendering parts of software under test (SUT). Such test data can improve, input fuzzing. To assess the proposed approach, we developed a modular file format fuzzer, \iust. Our conducted experiments on the MuPDF, a lightweight and favorite portable document format (PDF) reader, reveal that \iust{} reaches high coverage of SUT in comparison with the state-of-the-art tools such as learn\&fuzz, AFL, Augmented-AFL and random fuzzing. We also observed that the simpler deep learning models, the higher code coverage.
\end{abstract}

\begin{keyword}
Test data generation, File format fuzzing, Code coverage, Neural language model, Recurrent neural network, Deep learning.
\end{keyword}

\end{frontmatter}

%\linenumbers

\section{Introduction}
\noindent Fuzzing \cite{Miller:1990:ESR:96267.96279,Miller1995, Forrester:2000:ESR:1267102.1267108, Miller:2006:ESR:1145735.1145743} is a dynamic software testing technique to detect faults and vulnerabilities in programs. To this aim, test data sets are generated and injected into the SUT as far as the program crashes, or an unexpected behavior is observed. File format fuzzing is of great significance in the case of software handling malformed and untrusted files, including web browsers, PDF readers and multimedia players \cite{Sutton:2007:FBF:1324770, Rathaus:2007:OSF:1536880}.

The primary challenge concerned with the file format fuzzer is to generate test data as files, covering execution paths of the SUT. To generate test data for fuzzing programs dealing with files as their major inputs, the fuzzer requires to know the file format. In fact, without prior knowledge of the file format, most of the generated test data may be rejected very soon after running the SUT, and this can result in low code coverage \cite{Pham:2016:MWF:2970276.2970316}. Manual extraction of a file format is a common solution to deal with this problem. However, such a solution is costly, time-consuming, and requires the file format specification which may not be available, always. Therefore, automatic detection of the file format has been of great concern to test data generation approaches \cite{Godefroid:2017:LML:3155562.3155573, Rawat2017VUzzerAE}.

\begin{figure}%[ht!]%[tbh!]%[ht]%[t!]
 	\centering
 	\includegraphics[width=0.6\textwidth, clip=true,  trim= 0 0 0 0]{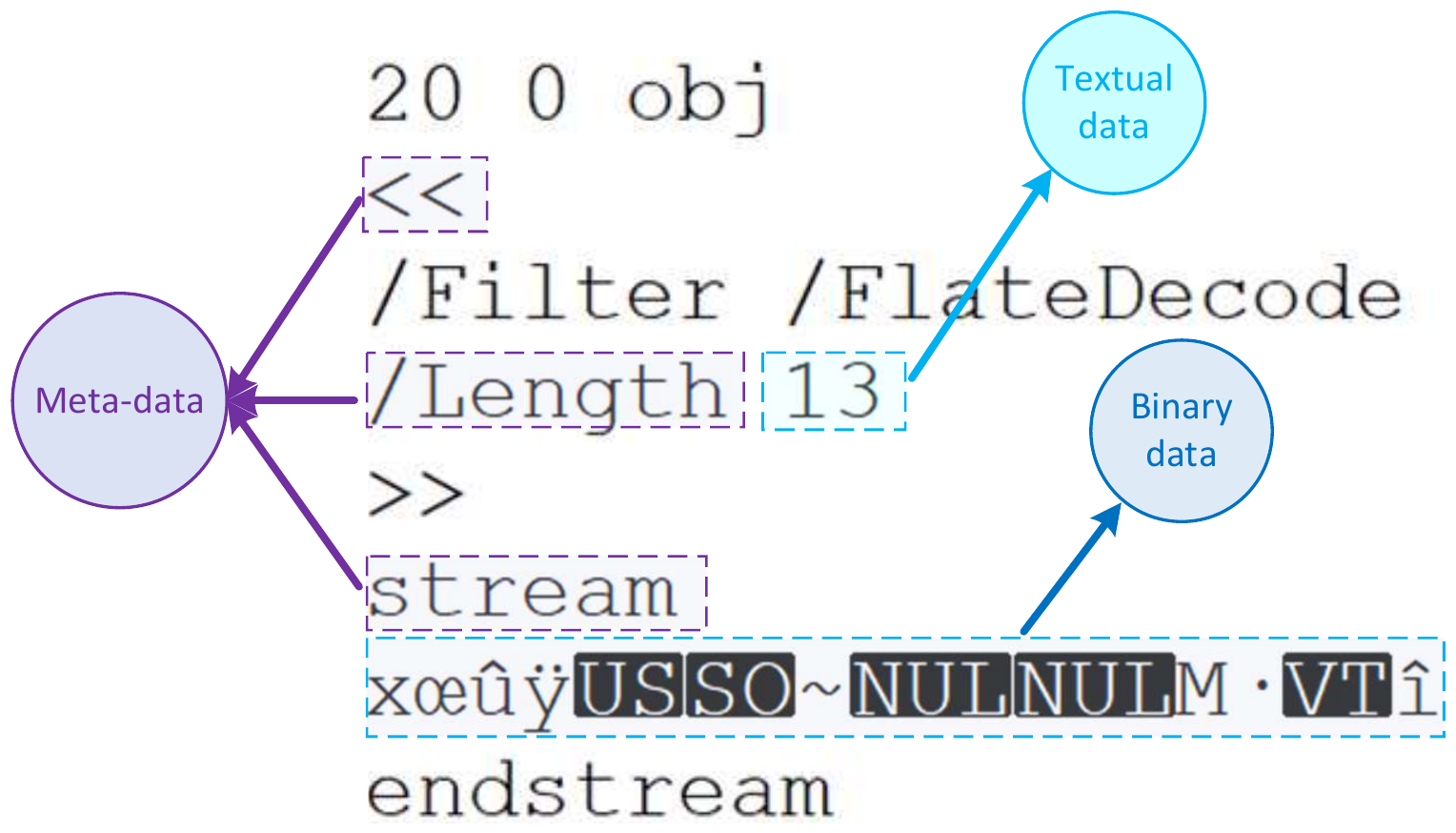}
 	\caption
 	{
 		A PDF data object and its different parts. An example of complex input structure.
 	}
 	\label{fig:complex-input-structure}
 \end{figure}

\begin{figure}%[ht!]%[tbh!]%[ht]%[t!]
	\centering
	\includegraphics[width=0.6\textwidth, clip=true,  trim= 0 0 0 0]{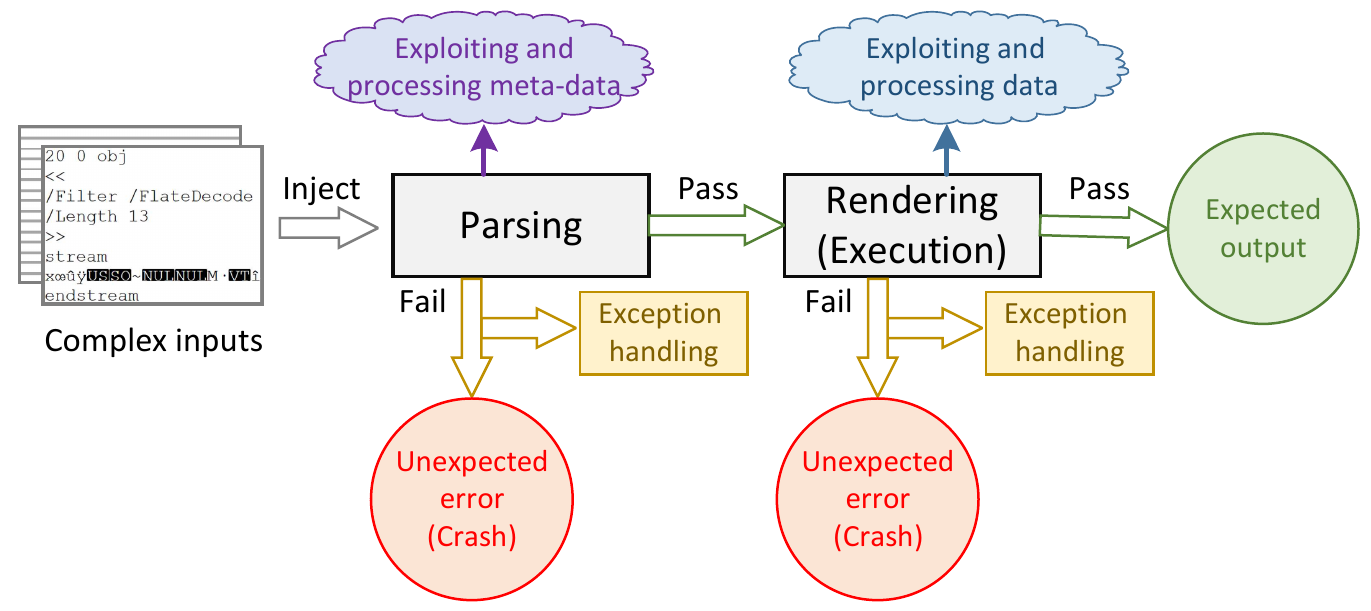}
	\caption
	{
		Stages of processing complex inputs in real-world applications.
	}
	\label{fig:complex-input-processing}
\end{figure}

An input with complex structure consists of textual and binary data fields described by some meta-data. File formats such as PDF \cite{Incorporated2006} are good example of complex input structures. Figure \ref{fig:complex-input-structure} shows a PDF data object and its different parts. A major dilemma concerned with the automatic detection of a file format is to distinguish meta-data such as tags and parameters, used to define the format, from the pure data, stored in the file. Typically, a program such as MuPDF \cite{MuPDF2018} processes a given PDF file in two distinct stages of \textit{parsing} and \textit{rendering} or \textit{execution} \cite{Skyfire2017, Pham:2016:MWF:2970276.2970316} shown in Figure \ref{fig:complex-input-processing}. Parsing mostly deals with meta-data. More specifically, the parser checks the format and at the same time loads the checked format into defined data structures in the main memory by employing meta-data parts. In the rendering stage, the loaded data is processed to generate the desired output. For example, the content of the file is shown to the user. 
 
 To perform a proper software testing and find more faults a file format fuzzer should fuzz the SUT both at the parsing and the rendering stages. Testing the parser requires all type of input including syntactically invalid test data. On the other hand, testing the rendering part requires a well-formed test data passed by the parser and meets the rendering code. Therefore, a test data generation method is needed to thoroughly understand the input structure and distinguish data from meta-data while generates new test data targeted at parsing or rendering stage. During fuzzing a program in addition to well-formed test data, malformed data are also required. The malformed data should be kept in an appropriate position within the input file such that defects of the corresponding SUT could be revealed \cite{Rawat2017VUzzerAE}. Here, the test data generation method is also responsible for generating malformed data in appropriate positions. 
 
 %Hence, a fuzzer must be able to distinguish between data and the meta-data kept in a file to determine its format.

There has been a significant challenge to achieve a relatively high code coverage in fuzzing. For instance, AFL \cite{Zalewsky2013} is a well-known mutation based file format fuzzer which follows an evolutionary approach, targeted at generating test data with maximum code coverage. AFL mutates a population of randomly selected files to achieve files, covering new paths which were not already observed. However, it is observed that for a large number of execution paths, AFL cannot meet an acceptable code coverage in a reasonable time and it is not suitable for fuzzing programs with complex input structures \cite{DBLP:journals/corr/abs-1711-04596}. 

A promising approach is to use learning techniques to generate test data. For example, Learn\&Fuzz \cite{Godefroid:2017:LML:3155562.3155573}, a generation based file format fuzzer, employs a sequence-to-sequence generative model \cite{NIPS2014_5346, DBLP:journals/corr/ChoMGBSB14}  to learn the structure of input files. The learned model is then used to generate files as input test data. Originally, the sequence-to-sequence model has intended for mapping two sequences of different domains \cite{NIPS2014_5346}. Learning the structure of a file is not, however, a mapping problem and can be done with simpler models. In Learn\&Fuzz \cite{Godefroid:2017:LML:3155562.3155573} only textual data are learned, while a complex file format contains both textual and non-textual data. Moreover, in the Learn\&Fuzz, generating data always begins with a fixed prefix \textit{“obj”} which results in a low variety of test data and finally, the presented fuzzing algorithm, called \textit{SampleFuzz} \cite{Godefroid:2017:LML:3155562.3155573}, may not terminate in all executions.

In this paper, to alleviate the aforementioned challenges, we propose a novel test data generation method to be applied to file format fuzzers. Our method learns the structure of complex input files by using the NLMs based on deep RNNs \cite{Mikolovinproceedings}, instead of sequence-to-sequence model. Two new fuzzing algorithms are introduced to fuzz both textual and binary parts of the input file, each of which targets one stage of processing the file. We also developed \iust, a new modular file format fuzzer to do fuzz testing. \iust{} can learn any complex file format, then generate and fuzz new test data fully automatic. In summary, our main contributions are as follows:

\begin{itemize}
	
	\item{We transfer the complex fuzzing test data generation problem into a language modeling task by applying NLM \cite{Mikolovinproceedings} to learn the structure of complex file formats and construct a generative model.}
	
	\item{We propose two specific fuzzing algorithms, \mdnf{} and \dnf, which are based on a learned generative model. The former targets the parsing stage and the latter focuses on rendering stage in the processing of the input files.}
	
	\item{We improve the coverage rate for SUTs with complex inputs by introducing a novel hybrid test data generation method which generates the textual data by generative model and binary data by mutation.}
	
	\item{We investigate the effectiveness of various language models with different configurations and several sampling strategies in the context of complex test data generation. Also, we study various parameters required when generating and fuzzing test data with deep learning techniques.}
	
	\item{We develop a modular file format fuzzer, \iust\footnote{The complete source code and documentation of \iust{} are available on our GitHub repository \href{https://github.com/m-zakeri/iust\_deep\_fuzz}{https://github.com/m-zakeri/iust\_deep\_fuzz}}, and make it publicly available to facilitate the other practitioners who work in dynamic software testing area as well as our dataset including a set of numerous PDF objects and PDF files, \iustpc\footnote{The \iustpc{} is available on our GitHub page: \href{https://github.com/m-zakeri/iust\_deep\_fuzz/tree/master/dataset}{https://github.com/m-zakeri/iust\_deep\_fuzz/tree/master/dataset}}., which was not previously available.
	}

	\item{We evaluate and compare the power of our presented method with different well-known fuzzers such as AFL \cite{Zalewsky2013}, Augmented-AFL \cite{DBLP:journals/corr/abs-1711-04596}, Learn\&Fuzz \cite{Godefroid:2017:LML:3155562.3155573}, and FileFuzz \cite{Sutton:2007:FBF:1324770}, regards to code coverage for the complex file format, i.e., PDF \cite{Incorporated2006} and real-world application, namely MuPDF \cite{MuPDF2018}.}

\end{itemize}

The proposed method is evaluated by learning PDF file format \cite{Incorporated2006} and then using the resulted format to generate PDF files as test data to fuzz an open-source PDF viewer, MuPDF \cite{MuPDF2018}. Our evaluation results indicate a relatively higher code coverage than the state of the art file format fuzzers \cite{Godefroid:2017:LML:3155562.3155573, Zalewsky2013}. Moreover, in this paper, we show that NLMs outperforms Learn\&Fuzz sequence-to-sequence model regarding the accuracy of learning file formats.

The rest of this paper is organized as follows. In Section 2, we briefly introduce language models (LMs) and RNNs as a fundamental concept used in our proposed method. In Section 3, we describe our proposed method for learning the structure of the file, generating, and fuzzing new test data. Section 4, deals with various experiments and evaluations, provided by applying our method in comparison with existing methods. Related works are discussed in Section 5. Finally, in Section 6, we conclude our proposed method and discuss some future works on fuzzing and complex test data generation.

\section{Language Model and Recurrent Neural Network}

\noindent We have applied language model to learn the structure of a file as a sequence of symbols. Language model is a fundamental concept in NLP, which allows predicting the next symbol in a sequence \cite{Jurafsky2017}. More precisely, LM is a probabilistic distribution over a sequence of words/symbols that identifies the probability of a given sequence. By using an LM, we can choose a more likely sequence among some existing ones. An LM for sequence $x=<x^{(1)},...,x^{(n)}>$ is defined as follows \cite{Luong2016}:

\begin{equation}\label{eq:language-model-formula}
p(x)=\prod_{t=1}^{n}p(x^{(t)}|x^{<t})
\end{equation} 

In Equation \ref{eq:language-model-formula}, each of the individual terms $p(x^{(t) }| x^{<t})$ indicates the conditional probability
of the current symbol $x^{(t)}$ given previous symbols $x^{<t}$, also referred to as the context. In practice, calculating this probability in the form of Equation \ref{eq:language-model-formula} is almost impossible, because we need to see all possible sequences. To overcome the computational challenge, traditional \textit{n-gram} LMs consider only a fixed context window of $n-1$ words, based on some sort of Markovian assumption. Although promising, in many cases, these models are not suitable for long sequences (more than 4 or 5 symbols) or unseen sequences \cite{Luong2016}. 

To address the n-grams problems, one can use a family of deep neural networks, namely, recurrent neural network for building an LM which is called neural language model \cite{Mikolovinproceedings}. NLMs can extend to longer context without encountering the zero probabilities problem. RNN is used to process sequential data. It handles the input sequence in a series of time steps and updates its memory to produce a hidden state, $h(i)$. Figure \ref{fig:example-of-rnn} shows a simple RNN with one hidden layer. In each time step $t$, one vector of the input sequence is processed. The feed forward equations of an RNN are defined as Equation \ref{eq:rnn-formula1} to \ref{eq:rnn-formula4} \cite{Goodfellow-et-al-2016}:

\begin{figure}%[ht!]%[tbh!]%[ht]%[t!]
	\centering
	\includegraphics[width=0.6\textwidth, clip=true,  trim= 0 0 0 0]{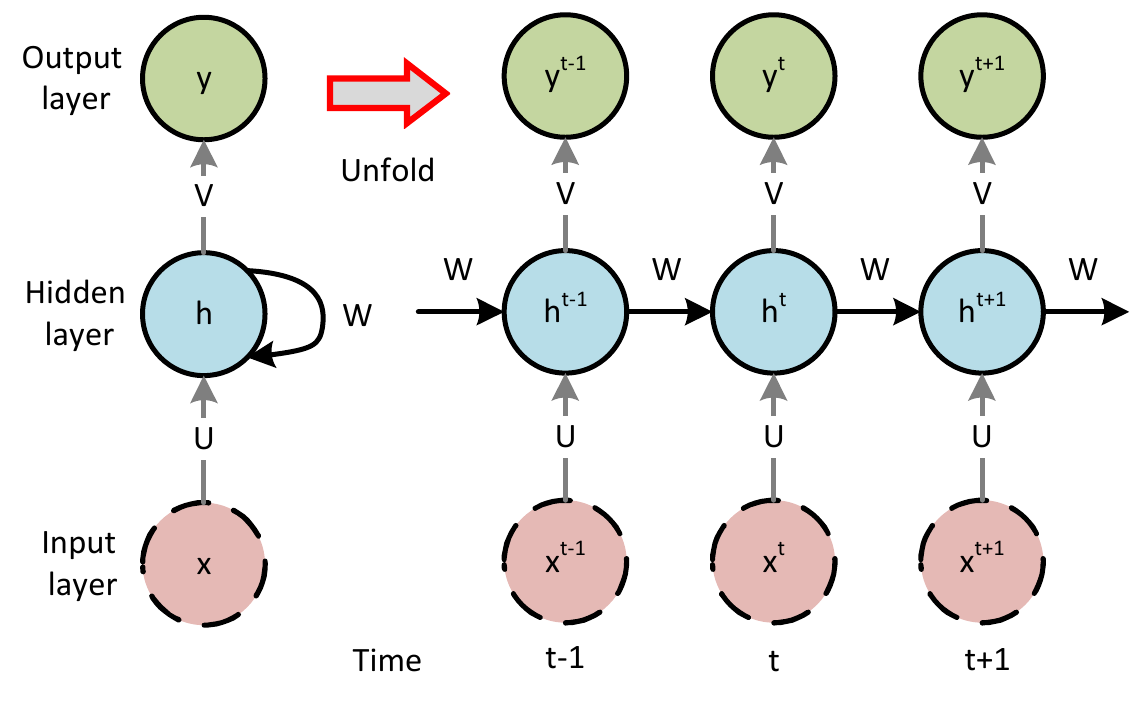}
	\caption
	{
		Computational graph of a recurrent neural network with one hidden layer \cite{Goodfellow-et-al-2016}.
	}
	\label{fig:example-of-rnn}
	
\end{figure}

\begin{equation}\label{eq:rnn-formula1}
z^{(t)}=Ux^{(t)}+Wh^{(t-1)}+b
\end{equation}
\begin{equation}\label{eq:rnn-formula2}
h^{(t)}=\sigma(z^{(t)})
\end{equation}
\begin{equation}\label{eq:rnn-formula3}
y^{(t)} = Vh^{(t)}+c
\end{equation}
\begin{equation}\label{eq:rnn-formula4}
\hat{y}^{(t)}=softmax(y^{(t)})
\end{equation}

\noindent where $b$ and $c$ are bias vectors and matrices $U$, $V$, and $W$, respectively, are the weights of input-to-hidden, hidden-to-output and hidden-to-hidden connections learned during the training of the network. Learning is achieved via defining a \textit{loss} (objective) function and using an optimization method to minimize it. $\sigma$ is an activation function such as \textit{sigmoid}. The \textit{softmax} function is applied on the output layer to convert the network output into a valid probability distribution.

LMs, as generative models, provide a probabilistic distribution over a sequence of symbols. By sampling from such a distribution, new sequences can be generated. In our proposed approach, each file is considered as a sequence of bytes, derived from the language of that file. Hence, we will build the corresponding language model for each file format. 

\section{Neural Fuzzing}
\begin{figure}
	\centering
	\includegraphics[width=0.95\textwidth, clip=true,  trim= 0 0 0 0]{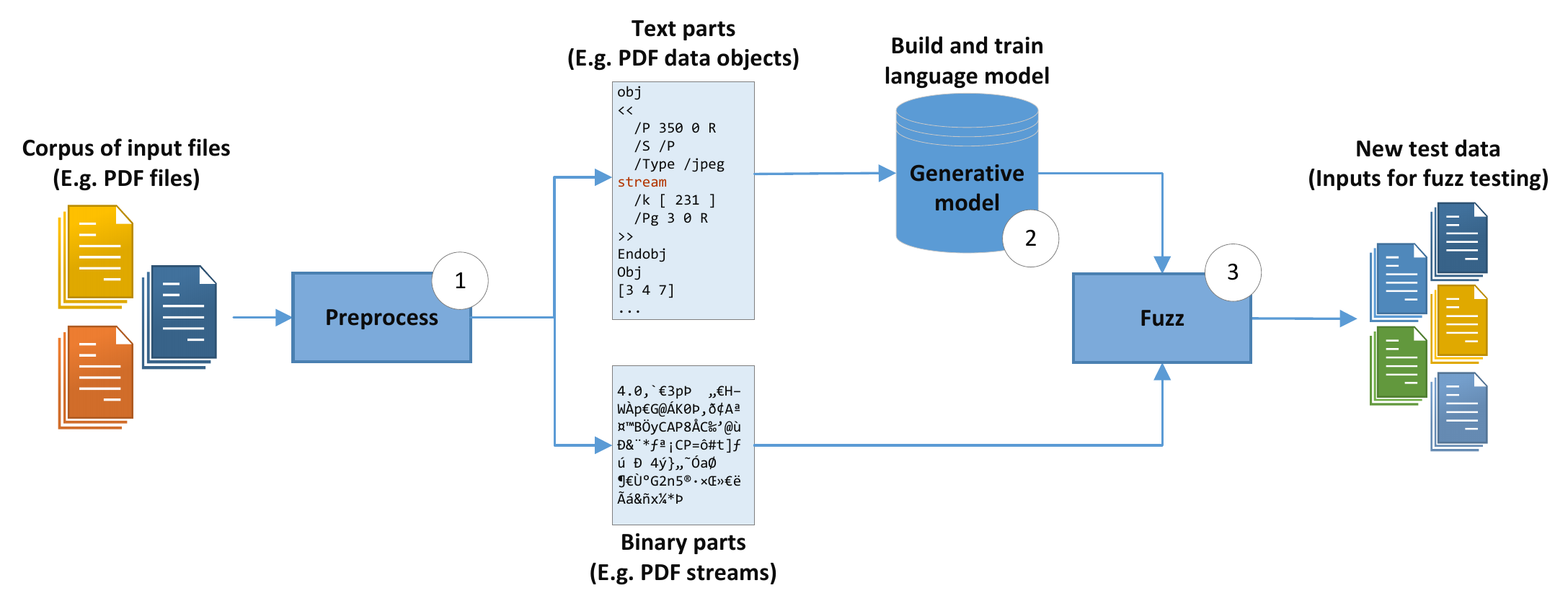}
	\caption
	{
		Flowchart of our hybrid test data generation method.
	}
	\label{fig:proposed-method}
	
\end{figure}

\noindent Our proposed method for test data generation consists of three main steps. First, gathering some sample data, i.e., input files, and preprocessing them. Second, training a language model on the provided train set. Third, generating and fuzzing test data through the learned model. Once the test data is generated, we can start fuzz testing on any given target. Figure \ref{fig:proposed-method} shows the flowchart of our proposed method, to be discussed in more details in the following sections.

\subsection{Overview}
\noindent As shown in Figure \ref{fig:proposed-method}, at the beginning (step 1), we collect a large number of sample files in the format that we want to learn, e.g., HTML or PDF files. Then, the binary (not-ASCII) parts within each sample file are replaced with a unique token, called binary token (BT). For example, we substitute all streams in a PDF file with the token \textit{stream}. By using this simple strategy, we can train an LM only with a set of ASCII sequences. Such a model probably predicts BT in the generation phase, and thus we can replace BT with a mutated binary part, which is previously generated by a mutation-based method. Of course, we need to keep the original binary elements for the future mutations and replacements. In this way, unlike the current methods which ignore the binary parts \cite{Godefroid:2017:LML:3155562.3155573}, we could generate both the binary and text parts of a test data, at once.

In the preprocessing phase, we also add an end token (ET) to the end of each file, indicating the completeness of the processed file. Then, we concatenate all of the files together and build a large sequence of files. This sequence is used for training LMs, but before training, we divide it into three separate sets as the train set, the test set, and the validation set. Such a division is required to measure model accuracy and perplexity on the unseen data. It, also, helps us in generating new data (See Section \ref{sec:generating-new-test-data}). Some file formats have ET explicitly. For example, HTML files are ended with the token \textit{</html>}. In such a case, there is no need to add an extra token. Now, we can define our models and train them on the provided dataset, (step 2).

Finally, our two newly introduced fuzzing algorithms are used to generate and fuzz the new test data, called \textit{DataNeuralFuzz} and \textit{MetadataNeuralFuzz} (step 3). The former is used to fuzz the data in a test data, and the latter is used to fuzz the format or meta-data of the file.

In order to study the effect of model complexity in learning file structures and using the resultant structure to generate test data, we build four models with different hyper-parameters and architectures based on the RNNs, shown in Table \ref{table:proposed-models}. In the first glance, it seems that the more complex the model, the more accurate language model describing the desired file format will be. However, our experiments have shown that this is not always true. Our experiments with the application of language models resulted from models of different complexities showed that in contrast, the simpler models resulted in language models that could reach relatively higher code coverage. 

Each one of the models, addressed in Table \ref{table:proposed-models}, uses the Long-Short Term Memory (LSTM) cells \cite{doi:10.1162/neco.1997.9.8.1735} as RNNs units which can learn long sequences of inputs. The first three models are unidirectional many-to-one LSTMs \cite{Karpathy2014}, whose architectures are similar to Figure \ref{fig:example-of-rnn}. These models are different w.r.t. the size of the hidden layers and units in each layer that affects the number of training parameters for each model. The last model (model 4) is a bidirectional LSTM. A bidirectional LSTM visits the input sequence in the backward and forward order. The bidirectional LSTM is composed of two unidirectional LSTMs. One of them processes the input sequence from left to right and the other from right to left. As a result, each forward pass has two outputs. A merge function is required to combine these outputs and produce a single output. We chose to use the sum function, which adds two output vectors elementwise.

\begin{table}[]
	\centering
	\caption{The proposed NLMs details including the number of trainable parameters.}
	\label{table:proposed-models}
	\resizebox{0.95\textwidth}{!}{%
		\begin{tabular}{@{}lllll@{}}
			\toprule
			\multicolumn{1}{c}{\begin{tabular}[c]{@{}c@{}}Model \\ ID\end{tabular}} & \multicolumn{1}{c}{Model name}                                               & \multicolumn{1}{c}{\begin{tabular}[c]{@{}c@{}}Number of \\ hidden layer(s)\end{tabular}} & \multicolumn{1}{c}{\begin{tabular}[c]{@{}c@{}}Number of units\\ in each layer\end{tabular}} & \multicolumn{1}{c}{\begin{tabular}[c]{@{}c@{}}Number of\\ trainable parameters\end{tabular}} \\ \midrule
			1                                                                       & \begin{tabular}[c]{@{}l@{}}Unidirectional LSTM\\  (Many to One)\end{tabular} & 1                                                                                        & 128                                                                                         & 127584                                                                                       \\
			2                                                                       & \begin{tabular}[c]{@{}l@{}}Unidirectional LSTM \\ (Many to One)\end{tabular} & 2                                                                                        & 128                                                                                         & 238656                                                                                       \\
			3                                                                       & \begin{tabular}[c]{@{}l@{}}Unidirectional LSTM\\  (Many to One)\end{tabular} & 2                                                                                        & 256                                                                                         & 870464                                                                                       \\
			4                                                                       & \begin{tabular}[c]{@{}l@{}}Bidirectional LSTM \\ (Many to One)\end{tabular}  & 2                                                                                        & 128                                                                                         & 469056                                                                                       \\ \bottomrule
		\end{tabular}%
	}
\end{table}

\subsection{Training the Model}
\noindent The training process for all the models, shown in Table \ref{table:proposed-models}, is the same. Neural networks are trained in a supervised mode; that is, an output label is required for each input of the network. To train each model, we need to specify the input and output of the corresponding deep neural network. We split the train set sequence $S$ into multiple smaller subsequences with fixed length $d$ such that the $i^{th} $ subsequence, $x_i$, will be:

\begin{equation}
	x_i=S[i*j: (i*j)+d]  
\end{equation}

\noindent where $S[l:u]$ is a subsequence of $S$ between the indices $l$ and $u$ and $j$ is a jump step, indicating the forward jump to select the next subsequence from original sequence, $S$; $x_i$s are input sequences for the model. The corresponding output or in other words label to each input sequence, $x_i$, is defined as: 

\begin{equation}
Y_{x_i}=S[(i*j)+d+1]
\end{equation}

Indeed, the output is the next symbol of the input sequence. After generating all input sequences and their corresponding output symbols, the model can be trained. During training, the model learns the conditional probability 
$p(x^{(i+d+1)}|<x^{(i)},...,x^{(i+d)}>)$ 
which will eventually enables it to predict the occurrence of the next symbol, $x^{(i+d+1)}$, of the given subsequence, $x_i$. 

\begin{figure}
	\centering
	\includegraphics[width=0.95\textwidth, clip=true,  trim= 0 0 0 0]{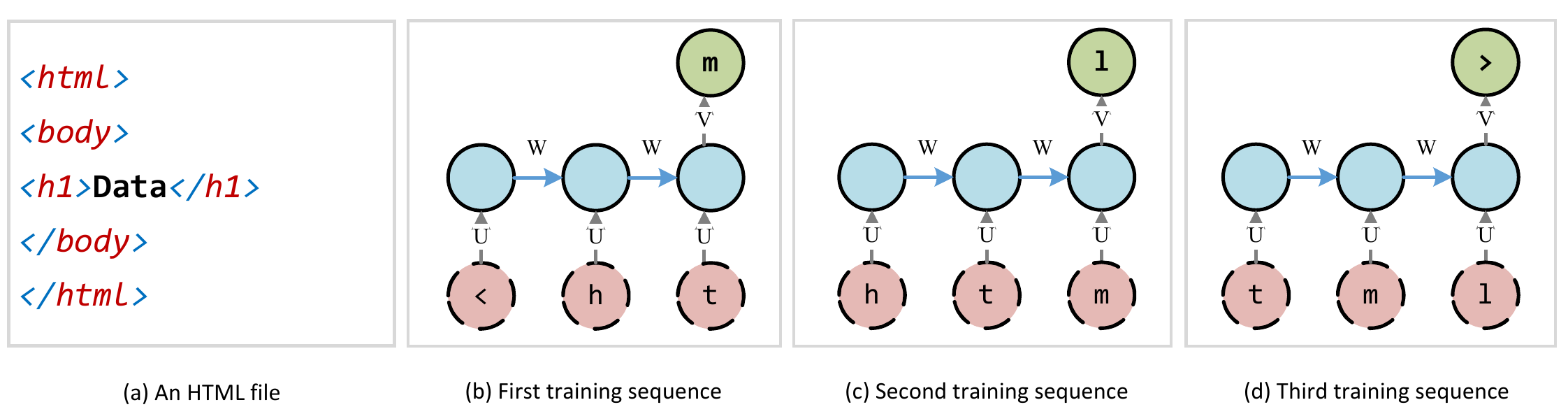}
	\caption
	{
		An example of training many to one RNN on a sample HTML file.
	}
	\label{fig:toy-example}
	
\end{figure}

Figure \ref{fig:toy-example} shows an example of the above-mentioned training method on a sample HTML file. The first three training sequences and their presentation to the network are shown in the figure. The parameters $d$ and $j$ are set to 3 and 1, respectively. In practice, $d$ can be set to a large number, i.e., $40$ or even $100$ which makes it possible to learn  long dependencies.

\subsection{Generating New Test Data}\label{sec:generating-new-test-data}
\noindent After the training process is completed, we can use the learned model to generate new data. To this aim, we first randomly select a prefix $P$ of length $d$, from the test set sequences, and feed it to the model. The model makes predictions for the next symbol as a valid distribution on all symbols. Afterward, a symbol form this distribution is selected and is augmented to the end of $P$. Next, the first symbol of $P$ is removed, while the length of $P$ is yet $d$. Now, we query the model with the updated prefix and generate the next symbol. This process will continue until the end token $ET$ is produced.

There are many strategies to choose one symbol from the output distribution. A straightforward strategy is to select the symbol with the highest probability, greedily. Such a strategy results in a well-formed file. However, the generated test data is always limited to the number of prefixes, i.e., the size of the test set. Another common strategy is to sample the predicted distribution as a multinomial distribution. The sampling can result in various test data, but it does not guarantee that they are all well-formed. Hence, we need a mechanism to control the diversity of generated test data during sampling.

In \cite{Godefroid:2017:LML:3155562.3155573} the authors have introduced \textit{SampleSpace} which combines a greedy selection with sampling, but the method is somewhat complicated and not a clear way to generate new test data. Their pure sampling strategy has produced better results. We have introduced a hyperparameter, diversity, to our sampling strategy. Diversity, $D$, is a real number in the interval $(0,1]$. At the generating phase, the model prediction values are divided by the diversity, $D$. After applying the $softmax$ function; sampling is done. As a result, a lower diversity causes the sampling strategy to close a greedy strategy and generates less various but well-formed test data. On the other hand, a higher diversity makes the sampling strategy to get away from the greedy strategy and creates more various yet mal-formed test data. 

\subsection{Fuzzing Test Data}
\noindent When we generate test data using the models in Table \ref{table:proposed-models} and the sampling strategy, outlined in the previous section, there will be an inherent variation in the generated data, and so these data can be used as test data. However, learning the file structure and fuzzing it are two ends of a spectrum. Learning wants to capture the structure of well-formed files and generates files which can pass the file parser while fuzzing intends to break-down the file structure, hope to make the program execution fail. In this section, we present the neural fuzzing algorithms with the aim of establishing a tradeoff between the two previous goals and the final generation of the test data for file format fuzzing. Our algorithms extend and improve the \textit{SampleFuzz} algorithm \cite{Godefroid:2017:LML:3155562.3155573}. As we mentioned, a file consists of two parts of the data and meta-data, each of which is processed in a separate stage. We introduce two algorithms called \textit{DataNeuralFuzz} to fuzz data, targeted at rendering stage and \textit{MetadataNeuralFuzz} to fuzz meta-data, aimed at parsing stage.

\textit{DataNeuralFuzz} is shown in Algorithm \ref{alg:data_neural_fuzz} and \textit{MetadataNeuralFuzz} is shown in Algorithm \ref{alg:metadata_neural_fuzz}. Both algorithms take as inputs the learned model $M$, sequence prefix $P$, diversity $D$, fuzzing rate $FR$, end token $ET$, binary token $BT$ and return as output the test data $TD$. Each algorithm has a main loop which continues until the $ET$ is not generated. Inside the while loop, $M$ is sampled with $D$. Then the predicted symbol is modified (fuzzed) under certain conditions which are different in the algorithms. After exit the while loop, the algorithm checks whether $TD$ contain $BT$ or not. If it includes $BT$, then $BT$ is replaced by the actual binary part which is fuzzed in the mutation based method, for example randomly. Remember that we already stored binary parts when separating them from the original dataset. Finally, $TD$ is returned by the algorithm.

One of the main features of both of these algorithms is that unlike \textit{SampleFuzz}, they always terminate. Before each of the algorithms goes into their while loop, a random integer number with the minimum value $a$ and the maximum value $b$ is set as the maximum length of $TD$, the $MaxLen$ variable. During generation of $TD$ in while loop if $ET$ is not produced by model and the length of $TD$ is larger than $MaxLen$, then $ET$ is added to the end of $TD$ by the algorithm, and the while loop will break. The values of $a$ and $b$ should be determined by the tester. A good practice is to set them around the average length of dataset files.
%%%%%%%%%%%%%%%%%%%%%%%%%%%%%%%%
%% neural fuzzing algorithms %%%
%%%%%%%%%%%%%%%%%%%%%%%%%%%%%%%%
\begin{algorithm}
			%\onehalfspacing
			\caption{DataNeuralFuzz} \label{alg:data_neural_fuzz}
				\DontPrintSemicolon
				\setcounter{AlgoLine}{0}
				\LinesNumbered
				
				\SetKwFunction{Random}{Random}
				\SetKwFunction{RandInt}{RandInt}
				\SetKwFunction{Predict}{Predict}
				\SetKwFunction{EndsWith}{EndsWith}
				\SetKwFunction{Sample}{Sample}
				\SetKwFunction{Chars}{Chars}
				\SetKwFunction{Len}{Len}
				\SetKwFunction{AddBinaryPart}{AddBinaryPart}
				\SetKwFunction{MutateBinaryPart}{MutateBinaryPart}
				\SetKwInput{KwData}{Input}
				\SetKwInput{KwResult}{Output}
				
				\KwData{Learned model $M$, Sequence prefix $P$, Diversity $D$, Fuzzing rate $FR$, End token $ET$, Binary token $BT$}
				\KwResult{Test data $TD$}
				\BlankLine
				$TD$  $\gets$ $P$\;
				
				$MaxLen$  $\gets$ \RandInt($a$, $b$)\;
				
				\While{$not$ \EndsWith($TD$, $ET$)}
				{
					$predicts$  $\gets$ \Predict($M$($P$))\;
					
					$c$, $p(c)$  $\gets$ \Sample($predicts$, $D$) \tcc*{Sample c from the learned model}\;
					
					$p\_fuzz$  $\gets$ \Random($0,1$) \tcc*{Decide whether to fuzz}\;
					
					\If{ $p(c) < \alpha \wedge c\not\in $ \Chars($BT \bigcup ET$) $ \wedge p\_fuzz < FR$}
					{
						$c$  $\gets$ $argmin_{c'}\{ p(c') \in predicts \}$ \tcc*{Fuzz c by c' where c' is the lowest likelihood}\;
					} 
					
					$TD$  $\gets$ $TD$ + $c$\;
					
					$P$  $\gets$ $P[1:]$ + $c$ \tcc*{Propagate fuzz to prefix}\;
					
					\If{ \Len($TD$) > $MaxLen$ }
					{
						$TD$  $\gets$ $TD$ + $ET$ \;
						
						\textbf{break}\;
					}
					
				}
				
				\If {$BT \in TD$}
				{
					\tcc{Binary data fuzzing:}\;
					
					$TD$ $\gets$ \AddBinaryPart($TD$)\;
					
					$TD$ $\gets$ \MutateBinaryPart($TD$)\;
				}
				
				\textbf{return} $TD$\;
\end{algorithm}

\begin{algorithm}
			%\onehalfspacing
			\caption{MetadataNeuralFuzz} \label{alg:metadata_neural_fuzz}
				\DontPrintSemicolon
				\setcounter{AlgoLine}{0}
				\LinesNumbered
				
				\SetKwFunction{Random}{Random}
				\SetKwFunction{RandInt}{RandInt}
				\SetKwFunction{Predict}{Predict}
				\SetKwFunction{EndsWith}{EndsWith}
				\SetKwFunction{Sample}{Sample}
				\SetKwFunction{Chars}{Chars}
				\SetKwFunction{Len}{Len}
				\SetKwFunction{AddBinaryPart}{AddBinaryPart}
				\SetKwFunction{MutateBinaryPart}{MutateBinaryPart}
				\SetKwInput{KwData}{Input}
				\SetKwInput{KwResult}{Output}
				
				\KwData{Learned model $M$, Sequence prefix $P$, Diversity $D$, Fuzzing rate $FR$, End token $ET$, Binary token $BT$}
				\KwResult{Test data $TD$}
				
				\BlankLine
				
				$TD$ $\gets$ $P$\;
				
				$MaxLen$  $\gets$ \RandInt($a$, $b$)\;
				
				\While{$not$ \EndsWith($TD$, $ET$)}
				{
					$predicts$  $\gets$ \Predict($M$($P$))\;
					
					$c$, $p(c)$  $\gets$ \Sample($predicts$, $D$) \tcc*{Sample c from the learned model}\;
					
					$p\_fuzz$  $\gets$ \Random($0,1$) \tcc*{Decide whether to fuzz}\;
					\tcc{Simpler condition to fuzz a symbol in meta-data mode:}
					\HiLi \If{$ p(c) > \beta \wedge p\_fuzz < FR  $}
					
					{
						
						\HiLi $c'$  $\gets$ $argmin_{c"}\{ p(c") \in predicts \}$ \tcc*{Fuzz c by c' where c' is the lowest likelihood}\;
					} 
					
					\HiLi $TD$  $\gets$ $TD$ + $c'$\;
					
					$P$  $\gets$ $P[1:]$ + $c$\; \tcc*{Don't propagate fuzz to prefix}
					
					\If{ \Len($TD$) > $MaxLen$ }
					{
						$TD$  $\gets$ $TD$ + $ET$ \;
						
						\textbf{break}\;
					}
					
				}							

				\If {$BT \in TD$}
				{
					\tcc{Binary data fuzzing:}\;
					
					$TD$  $\gets$ \AddBinaryPart($TD$)\;
					
					$TD$  $\gets$ \MutateBinaryPart($TD$)\;
				}
				
				\textbf{return} $TD$\;
	\end{algorithm}
%%%%%%%%%%%%%%%%%%%%%%%%%%%%%%%%
%% end of fuzzing algorithms %%%
%%%%%%%%%%%%%%%%%%%%%%%%%%%%%%%%

\subsubsection{DataNeuralFuzz}
\noindent The \textit{DataNeuralFuzz} algorithm is aimed at fuzzing data stored in a file. The most apparent property of the data stored in different files is its high variety. Therefore, it is observed that the learned model predicts the lower probability for the stored data than the meta-data, describing the files format. This means the model prediction vector in the positions containing pure data is smoother than the positions containing meta-data. Using this property of the probabilities stored in the prediction vector, the type of the data, as pure or meta-data, could be determined. To determine the data type, as pure or meta-data, we set a threshold, $\alpha$, obtained by experiment as the borderline. If the probability of a symbol $c$, i.e., $p(c)$, is less than $\alpha$, then $c$ is considered as pure data. \textit{DataNeuralFuzz} replaces the pure data item $c$ with $c'$, where $c'$ has the lowest likelihood, provided that:
\begin{enumerate}
\item{The probability $p(c)$, of the symbol $c$ is less than a given threshold, $\alpha$.}

\item{ Symbol $c$ belongs to neither a $BT$ nor $ET$.}

\item {The fuzzing rate, $FR$, given by the tester, is higher than a random number, $p\_Fuzz$, which is generated by the i.i.d.\footnote{Independent and identically distributed} random generator.} 
\end{enumerate}

The fuzzing rate, $FR$, indicates the percentage of data, to be fuzzed during test data generation. For example, if $FR$ is set to $0.1$, only 10 percent of data will be fuzzed by the algorithm. Also, we are not willing to fuzz critical token, i.e., $BT$ and $ET$ because these tokens are inserted into the files to address the binary sections and end of the file, respectively. That is why, it is ensured that the pure data item, $c$, does not belong to $BT$ and $ET$. 

Another aspect of pure data is that it appears as tokens with a length longer than one. Our \textit{DataNeuralFuzz} algorithm is designed to fuzz a pure data token by changing one or more symbols of the token. It is suggested \cite{Takanen:2008:FSS:1404500, Rathaus:2007:OSF:1536880} to change a data token with the highest possible value, dependent on the type of the token. Experimentally, the best practice in fuzzing is to replace a data token with its boundary values. For example, it is a good idea to use $999...9$ instead of an integer data. In general, it is widely known that boundary values, used as input data, may result in a crash in the rendering stage of the SUT execution.

The learned model can be used to generate any file, as input for fuzzing the SUT. To do so, a file, as an input string of fixed length, is given to the model, and the model generates the next symbol. Next time the input string is shifted one symbol ahead and this time the input string will include the newly generated symbol. The resultant string is again fed to the learned model to generate the second symbol. This process is repeated as far as enough symbols are created and a new input file is built. In order to increase the effectiveness of the generated input files, each time a new symbol is generated by the learned model we fuzz the symbol before using it, provided that the above mentioned three conditions are held. Each time that the model decides to fuzz the first symbol of a data token by adding this symbol to next prefix, we let the model stay in fuzzy prediction state. Next time that the learned model wants to predict a symbol, its prediction will be affected by the fuzzed symbol which probably results in another malformed symbol. We call this mechanism \textit{"propagating fuzz to the prefix"}.

\subsubsection{MetadataNeuralFuzz}
\noindent As described above our fuzing algorithm consists of two distinct parts, \textit{DataNeuralFuzz} and \textit{MetadataNeuralFuzz}, to generate and fuzz the pure data and format/meta-data of the generated file, respectively.  In fact, the generated file is further malformed to achieve a higher probability of making the SUT execution to crash. \textit{MetadataNeuralFuzz} attempts to crash the file format parser of the SUT. To avoid being trapped by exception handling mechanism used in the SUT parsing stage, \textit{MetadataNeuralFuzz} attempts to:

\begin{enumerate}
	\item{Applies the learned model, describing the appropriate structure of the files, to generate a new file, to test the SUT.}
	\item{Fuzz some of the symbols, describing the file format, with a certain percentage given by the tester.
	}
\end{enumerate}

\textit{MetadataNeuralFuzz} algorithm is intended to fuzz the file format while preserving the overall file structure as much as possible. In this way, \textit{MetadataNeuralFuzz} can check the parser robustness against invalid or malformed file formats. The learned model by itself does not have any assumptions about meta-data and pure data. It simply predicts the probability of occurrence of the next symbol while generating a file. \textit{MetadataNeuralFuzz} fuzzes meta-data while generating it. To distinguish meta-data from pure data, \textit{MetadataNeuralFuzz} uses the frequency of the symbols, gained at the training step. In general, meta-data is repeated more than pure data in the corpus. It is observed that the learned model predicts meta-data with a higher probability, very close to one, than the pure data. If the probability of a predicted symbol, $c$, is more than a given threshold, $\beta$, the algorithm guesses that the symbol, $c$, probably belongs to file format and replaces it with a symbol of the lowest occurrence probability. In order to control the percentage of the fuzzed symbols, a fuzzing rate, $FR$, is used. \textit{MetadataNeuralFuzz} fuzzes meta-data provided that a randomly generated number $p\_fuzz$ is less than a predetermined fuzzing rate, $FR$, given by the tester.

\textit{MetadataNeuralFuzz} considers $ET$ and $BT$ for fuzzing because these tokens are parts of the format. When a symbol generated by the learned model is fuzzed, it is simply stored in the targeted file and does not affect the prediction of the next symbol by the learned model. In this way, it is ensured that the fuzzed symbol does not propagate to the next prefix (line 10 of the \textit{MetadataNeuralFuzz} algorithm). The differences between the two algorithms, \textit{MetadataNeuralFuzz} and \textit{DataNeuralFuzz}, are highlighted in the \textit{MetadataNeuralFuzz} algorithm which is shown in Algorithm \ref{alg:metadata_neural_fuzz}.

\subsection{Implementation}
\noindent To implement deep NLMs, we used a high level deep learning library, Keras \cite{chollet2015keras}. Keras includes a set of high-level APIs for building deep learning models written in Python and need a low-level runtime back-end to execute deep learning code. We have decided to use TensorFlow \cite{DBLP:journals/corr/AbadiABBCCCDDDG16}, a Google framework for machine learning tasks, as the back-end for Keras. We used cross-entropy as the objective function and Adam \cite{DBLP:journals/corr/KingmaB14} with learning rates $1\times10^{-4}$ and $1\times10^{-3}$ as optimizer algorithm in the training process. We also applied \textit{Dropout} \cite{JMLR:v15:srivastava14a} technique to prevent our models from overfitting.

The purpose of this paper is to provide a method for automatically generating test data. However, the test data generation alone is not enough for fuzz testing. For evaluating the proposed method, we need to have a file format fuzzer. The fuzzer injects test data to SUT and checks for unexpected results such as crash the memory of the SUT. We develop \iust{} as a modular file format fuzzer. \iust{} uses Microsoft Application Verifier \cite{ApplicationVerifier}, a free runtime monitoring tool, as a monitoring module to catch any memory corruption. It also uses VSPerfMon \cite{VSPerfMon}, another tool from Microsoft, to measure code coverage. 

\begin{figure}
	\centering
	\includegraphics[width=0.95\textwidth, clip=true,  trim= 0 0 0 0]{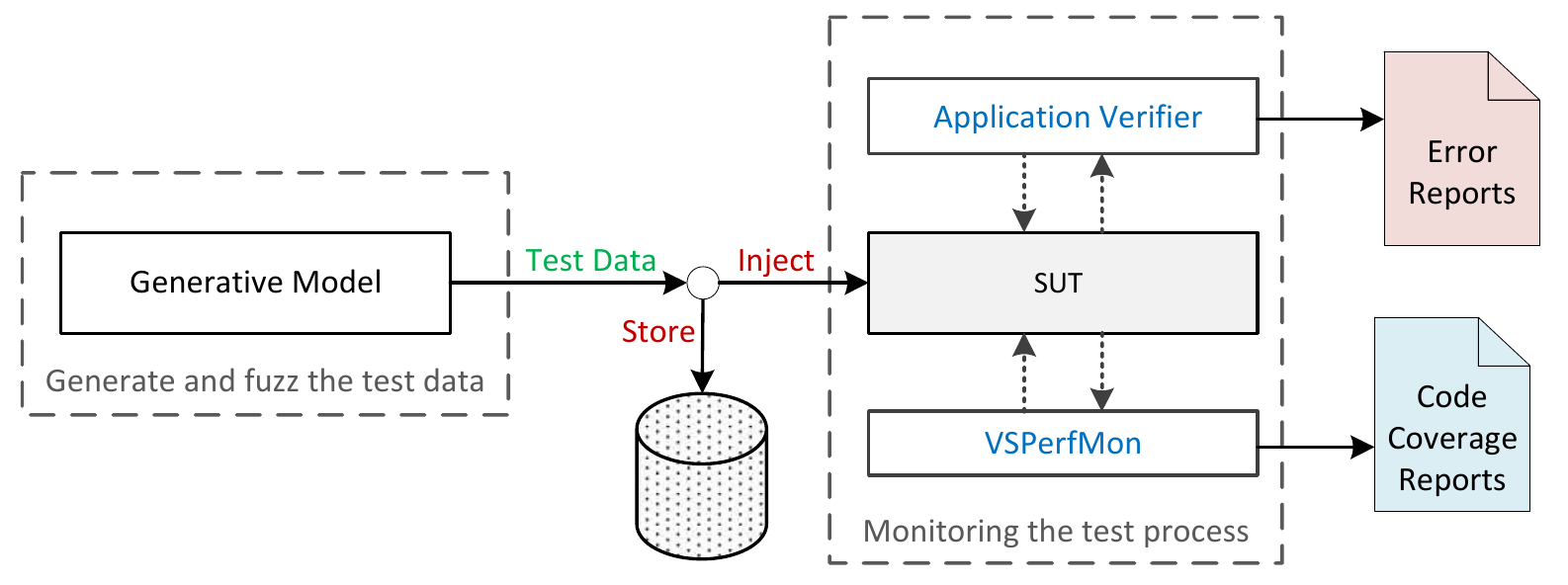}
	\caption
	{
		IUST DeepFuzz architecture.
	}
	\label{fig:iustdeepfuzz}
\end{figure}

The main module of \iust{} is a test data generator that implements our neural fuzz algorithms. These modules are connected using modest Python and batch scripts. \iust{} in the above configuration can run on the Windows operating system. To use it on the other operating systems we need to replace the monitoring tool, that is, Application Verifier \cite{ApplicationVerifier}. The test data generator is written in Python and could be run on any platform. The code coverage measurement module is only used for evaluating purposes, and our fuzz testing does not need it. \iust{} is a black box fuzzer \cite{Miller:1990:ESR:96267.96279} with hybrid test data generator. Each generated test data is stored on the disk before injection to SUT so if Application Verifier reports a crash, the test data which causes that crash can be retrieved to do fault localization process. Figure \ref{fig:iustdeepfuzz} shows the architecture and data flow of \iust.

\section{Experiments and Evaluations}
\noindent In this section, the results of our various experiments are presented. We use \iust{} to fuzz MuPDF \cite{MuPDF2018}, a free and an open source PDF, XPS, and E-book viewer, which takes as input complex PDF file \cite{Incorporated2006} and processes it. PDF is a complex and highly-structured file format. The full specification of the PDF file is described in Adobe PDF specification \cite{Incorporated2006}. Similarly, a brief description of the essential part of the PDF file specified in \cite{Godefroid:2017:LML:3155562.3155573}. The main part in the PDF file is data object which expresses all features and aspects of the file. Following the proposed method in \cite{Godefroid:2017:LML:3155562.3155573}, we trained our models on the set of PDF objects and then generated new PDF file to fuzz MuPDF viewer \cite{MuPDF2018}. 

We also implemented the Learn\&Fuzz method \cite{Godefroid:2017:LML:3155562.3155573} and evaluated it on MuPDF viewer \cite{MuPDF2018} because the Edge PDF parser and other material of Learn\&Fuzz, include dataset and model hyperparameters, did not make available publicly. In this way, we have been able to make a meaningful comparison between our proposed method and mentioned method as the most relevant work done in this area.

\subsection{Evaluation Metrics}
\noindent The primary purpose of fuzzing is to find faults and vulnerabilities in SUT which is directly associated with code coverage. The primary goal of learning file structure is to generate well-formed files which are associated with model accuracy. According to these facts, we consider the following metrics in our experiments to measure the effectiveness of our proposed method. 

\begin{enumerate}
	\item{\textbf{Models accuracy and error:} These metrics are based on objective function reported by Keras \cite{chollet2015keras} during training each model. Accuracy and error and are computed on the validation set data, which is derived from the dataset in the preprocessing phase.}
	
	\item{\textbf{Models perplexity:} Perplexity is the most common metric to evaluate an LM, and it is defined as \cite{mikolov2012statistical}:
	\begin{equation}\label{eq:ppl}
	\begin{split}
	PP_{LM}(x) & = \sqrt[n]{\prod_{i=1}^n(\frac{1}{p(x^{(i)}|<x^{(1)}, ..., x^{(i-1)}>)}} \\
	& = 2^{-\frac{1}{n}\sum_{i=1}^n\log_{2}{p(x^{(i)}|<x^{(1)}, ..., x^{(i-1)>})}}
	\end{split}
	\end{equation}
	In Equation \ref{eq:ppl}, $x$ is a sequence with length $n$ to evaluate the perplexity. The perplexity shows the difference between predicted sequence and test set sequence. So, the lower perplexity means the better LM. For each model, we compute perplexity on the validation set during training. We use perplexity to evaluate how proposed models are excellent in capturing the structure of the input file and to compare different proposed NLMs.
}

\item{\textbf{Code coverage:} For each test data execution, basic block coverage is measured by VSPerfMon tool \cite{VSPerfMon}. Basic block coverage is an extension of statement coverage, in which each sequence of non-branching statements is treated as one statement unit. The main advantage of basic block coverage is that it can be applied to object code with low overhead. Total coverage for test set is the union of individual coverages. VSPerfMon also reports line coverage which is the same statement coverage for the high-level code.}

\item{\textbf{Faults and vulnerabilities:} For each test data execution, Application Verifier \cite{ApplicationVerifier} creates a log file. We then search these log files with a simple script to find any error or security warning.}
	 
\end{enumerate}

The first two metrics, determine the effectiveness of learning file format and the next two metrics measure the quality and usefulness of fuzz testing.

\subsection{Experiments Setup}
\noindent Training the models in Table \ref{table:proposed-models} was performed on physical ubuntu 16.04 machine with single Nvidia GTX 1080 GPU, Intel Core i7 CPU and 20 gigabytes of RAM. Fuzz testing is done on virtual Windows 10 machine with Intel Core i7 CPU and 8 gigabytes of RAM. We used the final version of MuPDF viewer \cite{MuPDF2018} in time of doing our experiments, i.e., version MuPDF 2017-04-11\footnote{This release is available for download at \href{https://mupdf.com/release_history.html}{https://mupdf.com/release\_history.html}}. 

Before we can generate test data, we should train our models. Table \ref{table:model-runing-config} shows the critical hyperparameters of our models along with the number of epochs and training time for each model. The complexity of the model, i.e., the number of training parameters, increases with the model's ID. For more complex models, it is rational to have more training samples. Therefore, in models 3 and 4, we decrease the jump step which is lead to increase training samples. In model 3, we used the Dropout \cite{JMLR:v15:srivastava14a} with $p=0.3$ for regularizations purpose.  

\begin{table}[]
	\centering
	\caption{Training details for proposed models in Table \ref{table:proposed-models}.}
	\label{table:model-runing-config}
	\resizebox{0.85\textwidth}{!}{%
		\begin{tabular}{lllll}
			\hline
			\multicolumn{1}{c}{\multirow{2}{*}{Parameter}} & \multicolumn{4}{c}{Model ID}  \\ \cline{2-5} 
			\multicolumn{1}{c}{}                           & 1     & 2     & 3     & 4     \\ \hline
			Input sequence length (d)                      & 50    & 50    & 50    & 50    \\
			Jump step (j)                                  & 3     & 3     & 1     & 1     \\
			Number of training epochs                      & 50    & 50    & 50    & 50    \\
			One epoch training time (hour: min’)           & 1:00' & 1:45' & 5:30' & 9:30' \\
			Model size (megabytes)                         & 1.24  & 2.76  & 9.99  & 5.41  \\ \hline
		\end{tabular}%
	}
\end{table}

\subsection{Dataset and Host Files}
\noindent The successful training of deep neural networks requires a large and enough dataset. Hence, we collected a large corpus of PDF files from various source include Mozilla PDF.js open test corpus \cite{PDFjs}, some PDFs which are used in AFL \cite{Zalewsky2013} as initial seed and PDFs gathered from public web in different languages. Finally, we published \iustpc{} with more than $6,000$ PDF files. Such a corpus was not available publicly earlier, and it can also be used for others type of PDF manipulation and testing. 

To learn the statistical structure of the PDF objects, we extracted $500,000$ objects from \iustpc. About $27\%$ of these objects had a binary stream. We replaced binary streams with the binary token \textit{stream}, extracted and stored them into a separate dataset, and included modified objects in our training process. A key difference with \cite{Godefroid:2017:LML:3155562.3155573} is that we did not apply seed minimization before extracting objects because we want to learn the structure of the file and more data probably improve learning. The entire set of extracted PDF data objects is available beside the \iustpc.

Since we only learn and generate PDF objects, we need a mechanism to create complete PDF files. Follow the method presented in \cite{Godefroid:2017:LML:3155562.3155573} we decided to append newly generated objects to an existing well-formed PDF file, called \textit{host}. PDF files can be incrementally updated as described in the PDF Reference Guide \cite{Incorporated2006}. The new object appends to end of existing PDF, and its offset adds to the cross-reference table. This method allows one to update a PDF file without rewriting the entire file. Indeed, the new object rewrites the content of an existing object which is identified by an ID and absolutes the old one. More detail on the incremental update can be found on \cite{Incorporated2006}.

Next step is to choose a host file. In work proposed by \cite{Godefroid:2017:LML:3155562.3155573}, this is done almost randomly by only selecting the smallest three PDF files from their corpus. Against that work, in order to study the effect of the host complexity on the code coverage, we first compute code coverage for all of the PDF files in our corpus by running MuPDF and then select the three files with maximum, minimum, and average code coverage respectively as \textit{host1\_max}, \textit{host2\_min}, and \textit{host3\_avg}. 

\subsection{Baselines for Code Coverage}
\noindent To compare code coverage of newly generated PDF files with the existing PDF files, we first measured the MuPDF \cite{MuPDF2018} code coverage for every single host, then built $1,000$ PDF files with the objects that are selected from the test set, randomly. The objects are appended to the host files in two different modes:

\begin{enumerate}
	\item{\textbf{Single Object Update (SOU):} Find the last object ID in the host file and rewrite it with the new object. In this mode, only one object will be changed in each file.}
	
	\item{\textbf{Multiple Objects Update (MOU):} Rewrite the fix portion of the objects within each PDF file. In this mode first, the number of total objects in the host is computed, and then a randomly selected list of objects IDs will be overridden by new objects.}
\end{enumerate}

\begin{figure}
	\centering
	\includegraphics[width=0.95\textwidth, clip=true,  trim= 0 0 0 0]{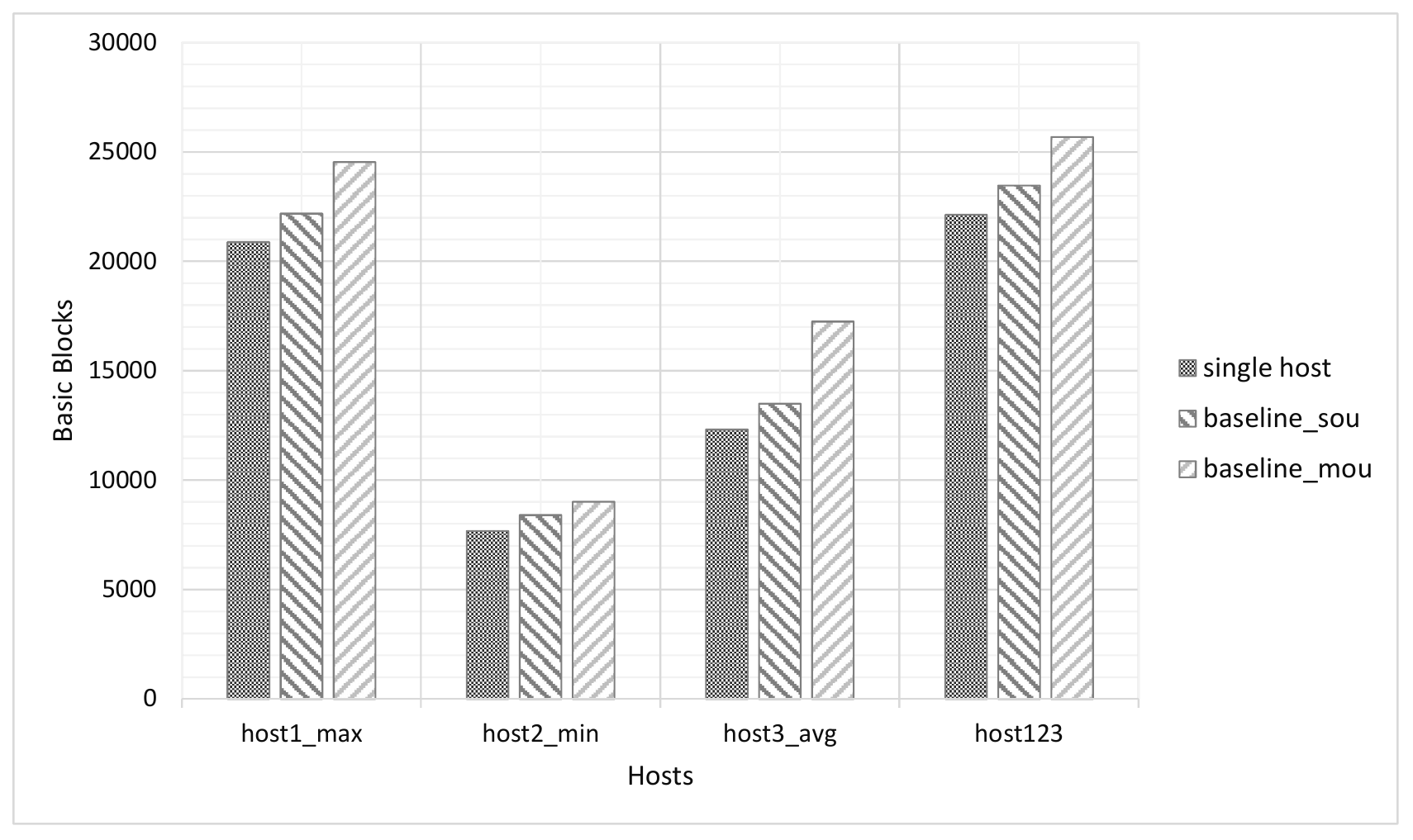}
	\caption
	{
		Code coverage of each host beside the baseline coverages.
	}
	\label{fig5:baseline-coverage}
	
\end{figure}

Table \ref{table:host-details} shows the number of objects in each host and the portion of rewrite objects in the MOU mode. Figure \ref{fig5:baseline-coverage} shows the code coverage obtained by running the MuPDF viewer on three hosts besides the coverage of two test suites one for SOU called \textit{baseline\_sou} and one for MOU called \textit{baseline\_mou}. \textit{host123} denotes the union of code coverages obtained from the hosts 1, 2 and 3. The following results are observed.

\begin{table}[]
	\centering
	\caption{Host files details include the number of objects exists in each host.}
	\label{table:host-details}
	\resizebox{0.85\textwidth}{!}{%
		\begin{tabular}{@{}lll@{}}
			\toprule
			Hosts      & Number of objects & A portion of rewrite objects in the MOU mode \\ \midrule
			host1\_max & 250               & 1/5                                          \\
			host2\_min & 9                 & 1/3                                          \\
			host3\_avg & 19                & 1/4                                          \\ \bottomrule
		\end{tabular}%
	}
\end{table}

\begin{itemize}
	\item {The code coverage of each baseline is higher than the coverage of the alone host. This means that changing the hosts leads to increase the code coverage.}
	
	\item{The baseline coverage has a direct relationship with host coverage. For example, \textit{host1\_max} has the highest code coverage among \textit{host1\_max}, \textit{host2\_min} and \textit{host3\_avg}. This shows that selecting an appropriate host file is an essential job and has a significant impact on the baseline coverage.}
	
	\item{Code coverage for \textit{baseline\_mou} is greater than \textit{baseline\_sou} in all cases. This incomes that further modification of the file content leads to increase the code coverage.
	}

	\item{
		The maximum code coverage belongs to \textit{host123} which shows every host has executed different basic blocks.
	}

	\item{
	Finally, the order of covered code is in the range 20,000 basic blocks which show MuPDF viewer \cite{MuPDF2018} is a large-size application and PDF files have a complex format. 
	}

\end{itemize}

\subsection{Model Evaluation}
\begin{figure}
	\centering
	\includegraphics[width=0.85\linewidth]{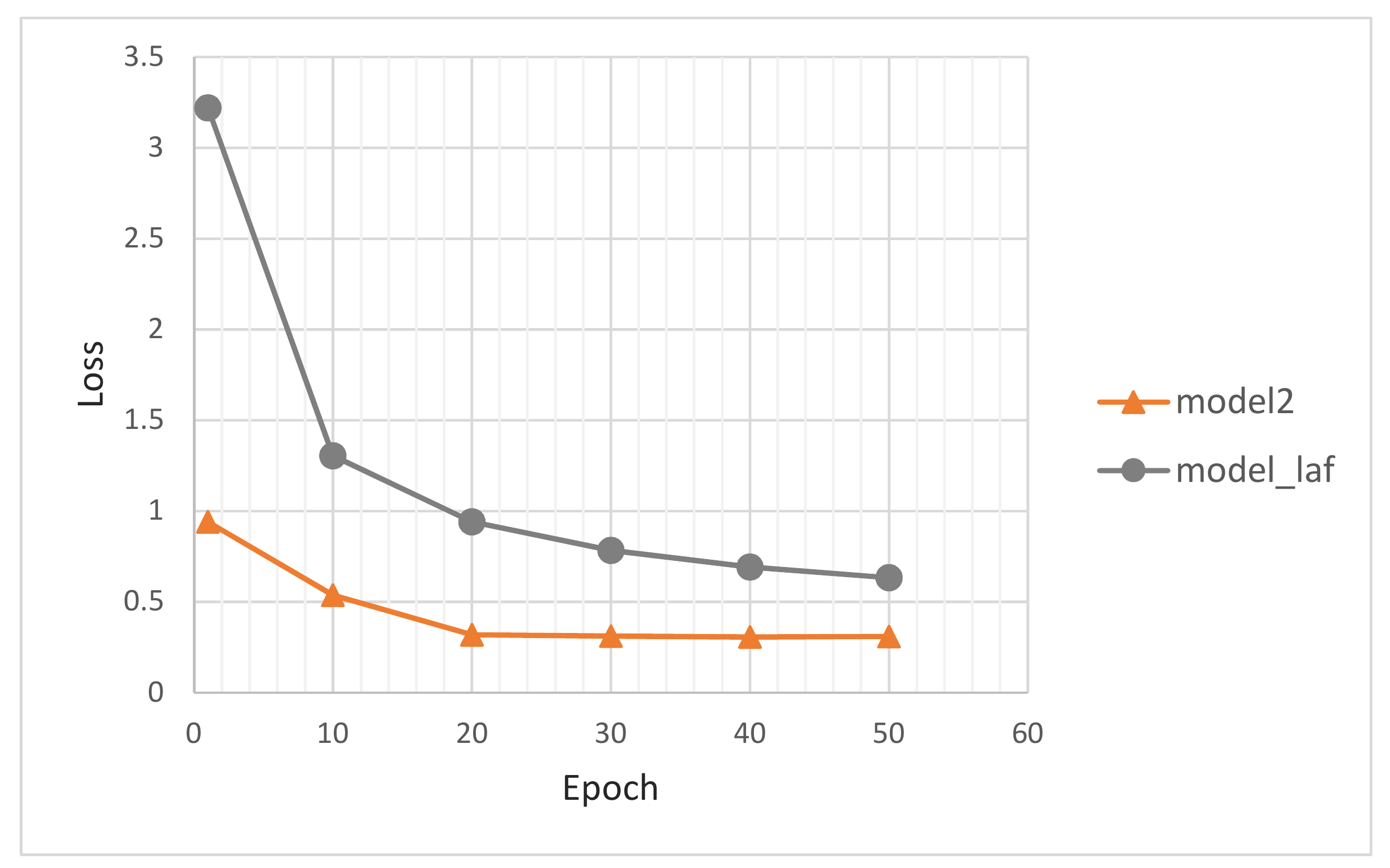}
	\caption{Validation error per epoch for model 2 and model \textit{laf} during the training process.}
	\label{fig:val-loss-model2-vs-laf}
\end{figure}

\noindent Table \ref{table:model-evaluation} shows the perplexity, accuracy, and error of our models after training for $50$ epochs. The last column named \textit{laf} shows this value for the Learn\&Fuzz model \cite{Godefroid:2017:LML:3155562.3155573}. These metrics are reported by Keras. The perplexity is computed by Equation \ref{eq:ppl}. Accuracy and error come from the cross-entropy loss function. Also Figure \ref{fig:val-loss-model2-vs-laf} shows the validation error diagram for model 2 and model \textit{laf} during the training process. The model 2 has been illustrated in this diagram because it is the most similar model to \textit{laf} in both architecture and hyperparameter setting. The following results are observed.

\begin{table}[]
	\centering
	\caption{Perplexity, accuracy, and error of proposed models. The best value in each row is bolded.}
	\label{table:model-evaluation}
	\resizebox{0.85\textwidth}{!}{%
		\begin{tabular}{@{}llllll@{}}
			\toprule
			\multicolumn{1}{c}{\multirow{2}{*}{Metric}} & \multicolumn{5}{c}{Model ID}                                \\ \cmidrule(l){2-6} 
			\multicolumn{1}{c}{}                        & 1     & 2     & 3              & 4              & laf       \\ \midrule
			Perplexity                                  & 1.440 & 1.391 & \textbf{1.335} & 1.350          & Undefined \\
			Maximum training accuracy                   & 0.886 & 0.902 & 0.893          & \textbf{0.909} & 0.820     \\
			Maximum validation accuracy                 & 0.884 & 0.895 & 0.904          & \textbf{0.905} & 0.800     \\
			Minimum training error                      & 0.353 & 0.298 & 0.324          & \textbf{0.276} & 0.623     \\
			Minimum validation error                    & 0.365 & 0.330 & \textbf{0.289} & 0.299          & 0.725     \\ \bottomrule
		\end{tabular}%
	}
\end{table}

\begin{itemize}
	\item{The error of all NLMs is less than \textit{laf} error, and their accuracy is more than it. This means that the NLM is better than the encoder-decoder model in the learning grammar of the file.}
	
	\item{The maximum accuracy belongs to model 4, our only bidirectional LSTM. This network processes the input sequence in both left-to-right and right-to-left direction. So it can reach an upper accuracy which results in a lower perplexity.}
	
	\item{In Figure \ref{fig:val-loss-model2-vs-laf}, the error diagram for model 2 always is under the model laf. Of course, the epochs have a different period, so a peer-to-peer comparison may not be exciting. However, we also see this relationship is true for an equal interval from the start of the training process.}
	
	\item{The maximum perplexity of all models, which is perplexity in the absence of an NLM, is 64 on our dataset. The perplexity after 50 training epochs is less than 1.5 which shows NLMs can learn the language of the file so excellent. The minimum perplexity belongs to model 3 that has the maximum number of trainable parameters.}
	
\end{itemize}

\subsection{Sampling Diversity and Code Coverage}
\noindent To study the impact of diversity on code coverage when generating test data, we produce 1,000 PDF files on each host, using the sampling strategy with different diversities 0.5, 1.0 and 1.5. This experiment provides information about the best model, host, diversity, and updating mode (i.e., SOU and MOU) in code coverage. As a result, we can choose the best configuration for use in fuzz testing. We save a checkpoint at the end of each epoch in training time then select the model with the minimum validation error between all checkpoints. We choose the best-learned model to sample it. 

\begin{figure}
	\centering
	\includegraphics[width=1.05\linewidth]{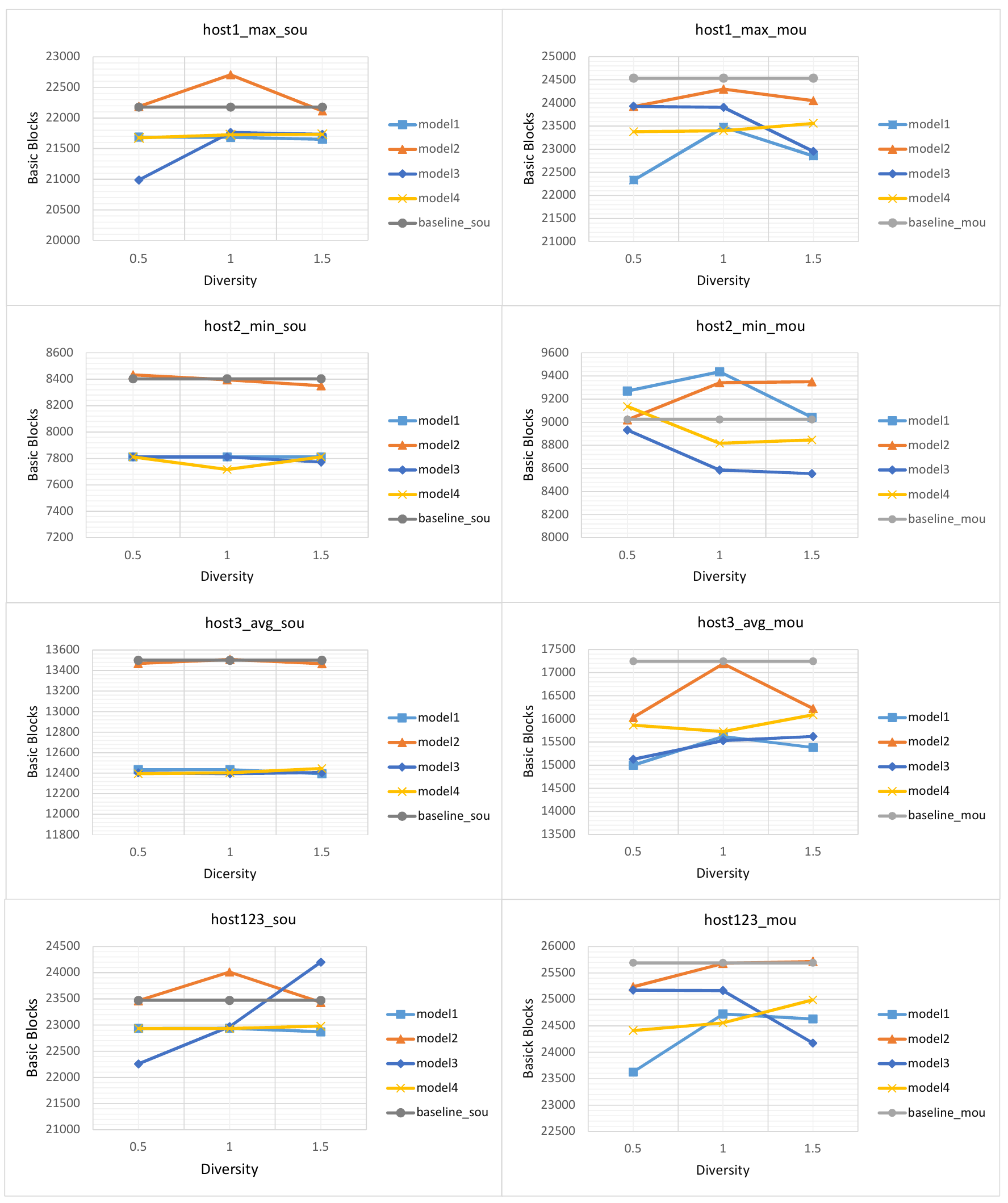}
	\caption{Code coverage per diversity for all models when evaluating them on each host file.}
	\label{fig:divs-all-model-host}
\end{figure}

Generating 1,000 PDF file with our models in SOU mode took about 60 minutes and in MOU mode took about 190 minutes. Also running each test suit on MuPDF viewer and obtaining the coverage took in average 65 minutes. In total, we generate and test 72,000 PDF files in this experiment. All code coverages are shown in figure \ref{fig:divs-all-model-host}. The following results are observed.

\begin{itemize}
	\item{Code coverage for generated data is less than the baselines code coverage in most cases because generated objects are not well-formed as real PDF objects in our test set. However, in such case, we see the increase in code coverage for example in diagram \textit{host2\_min\_mou}. This means that for small host adding the more content result in better code coverage.}
	
	\item{Increasing diversity leads to increase code coverage in Bidirectional LSTM (model 4) but not in the other models. In general, it seems that generating data with diversity one is more effective in most models and on most hosts w.r.t the code coverage of SUT. 
	}

	\item{Almost in all diagrams, model 2 outperforms other models in the code coverage. This means simpler NLMs act better than more complex NLMs.}
	
	\item{By looking at \textit{host123} diagrams, as an aggregation of the results, we can conclude that model 2 with diversity one is the best model for fuzz testing. Hence, we chose this model for using in our neural fuzz algorithms in Section \ref{sec:comparison-with-laf}.
	}
	
\end{itemize}

\subsection{Comparison with the Sequence-to-Sequence Model}\label{sec:comparison-with-laf}
\begin{figure}
	\centering
\begin{subfigure}
	{\linewidth}
	\centering
	\includegraphics[width=0.95\linewidth, clip=true,  trim= 0 0 0 0]{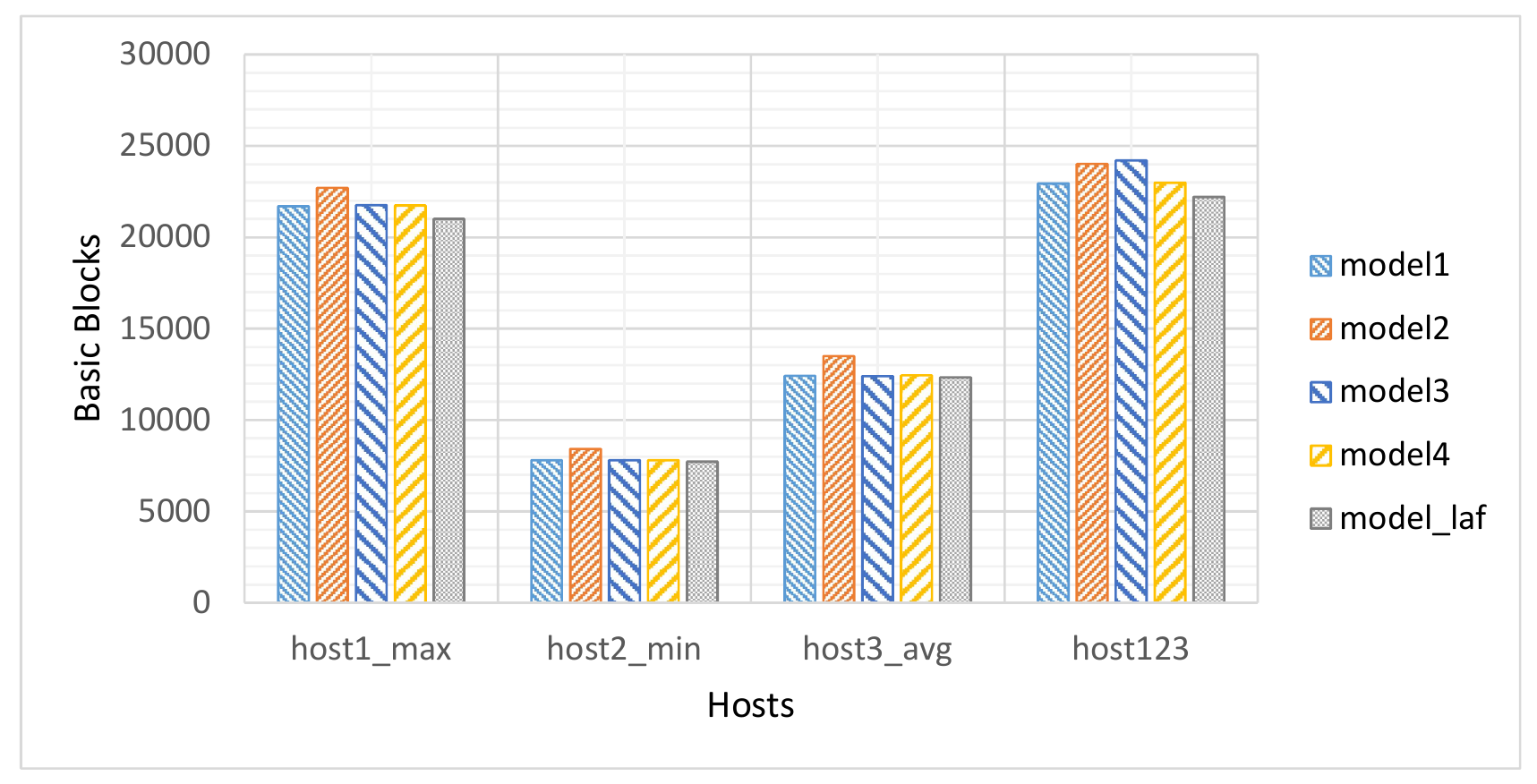}
	\caption{SOU mode}
	\label{fig8-cmp-laf-sou}
\end{subfigure}
%\vspace{0.75cm}
\begin{subfigure}
	{\linewidth}
	\centering
	\includegraphics[width=0.95\linewidth, clip=true,  trim= 0 0 0 0]{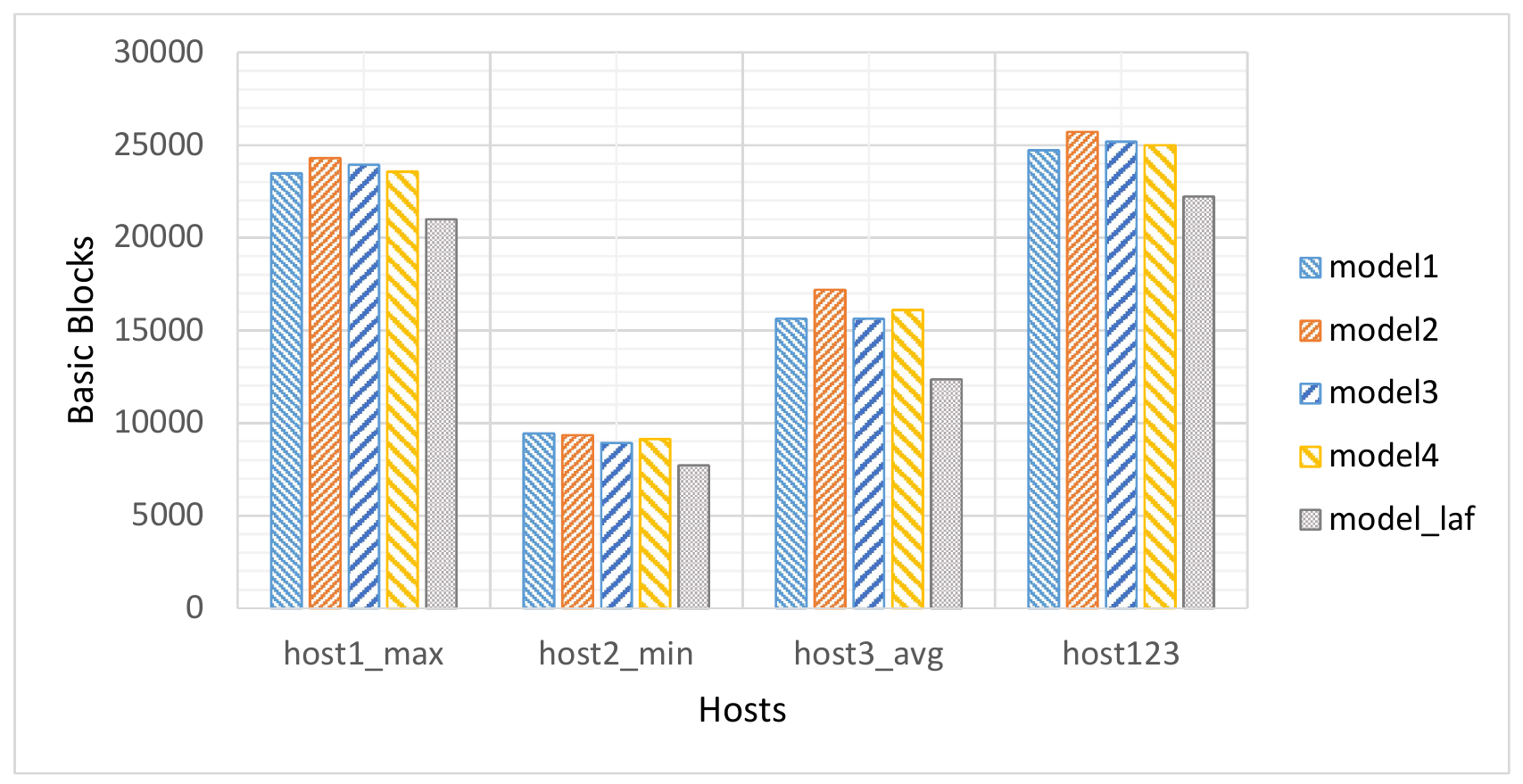}
	\caption{MOU mode}
	\label{fig8-cmp-laf-mou}
\end{subfigure}
	\caption{Code coverage per host for proposed models in comparison with the model \textit{laf} in SOU and MOU modes.}
	\label{fig:cmp-laf-sou-and-mou}
\end{figure}

\noindent To compare our models with the sequence to sequence model described in \cite{Godefroid:2017:LML:3155562.3155573}, we generate 1,000 PDF files with this model by using sampling strategy as the best strategy reported in \cite{Godefroid:2017:LML:3155562.3155573} and then obtain the code coverage for the created test suite. For each of our models, we choose the best code coverage from the previous section results. Figure \ref{fig:cmp-laf-sou-and-mou} shows the code coverage values in both MOU and SOU modes. We observed the following results.

\begin{itemize}
	\item{In SOU mode, all NLMs on \textit{host1\_max} have a better code coverage than \textit{laf}. However, on \textit{host2\_min} and \textit{host3\_avg} there is a subtle difference between models. }
	
	\item{
		In MOU mode, our proposed models are significantly better than the Learn\&Fuzz model. This show the effectiveness of MOU in generating PDF files. More changes result in higher code coverage. 
	}
	
	\item{In both modes, the increase in the code coverage of \textit{host1\_max} is higher than the increase in code coverage of \textit{host2\_min}. This means, in general, that more complex host files have more potential to execute new part of codes when their content is changed by the generative models.
	}
		
\end{itemize}

\begin{figure}
	\centering
	\includegraphics[width=0.85\linewidth]{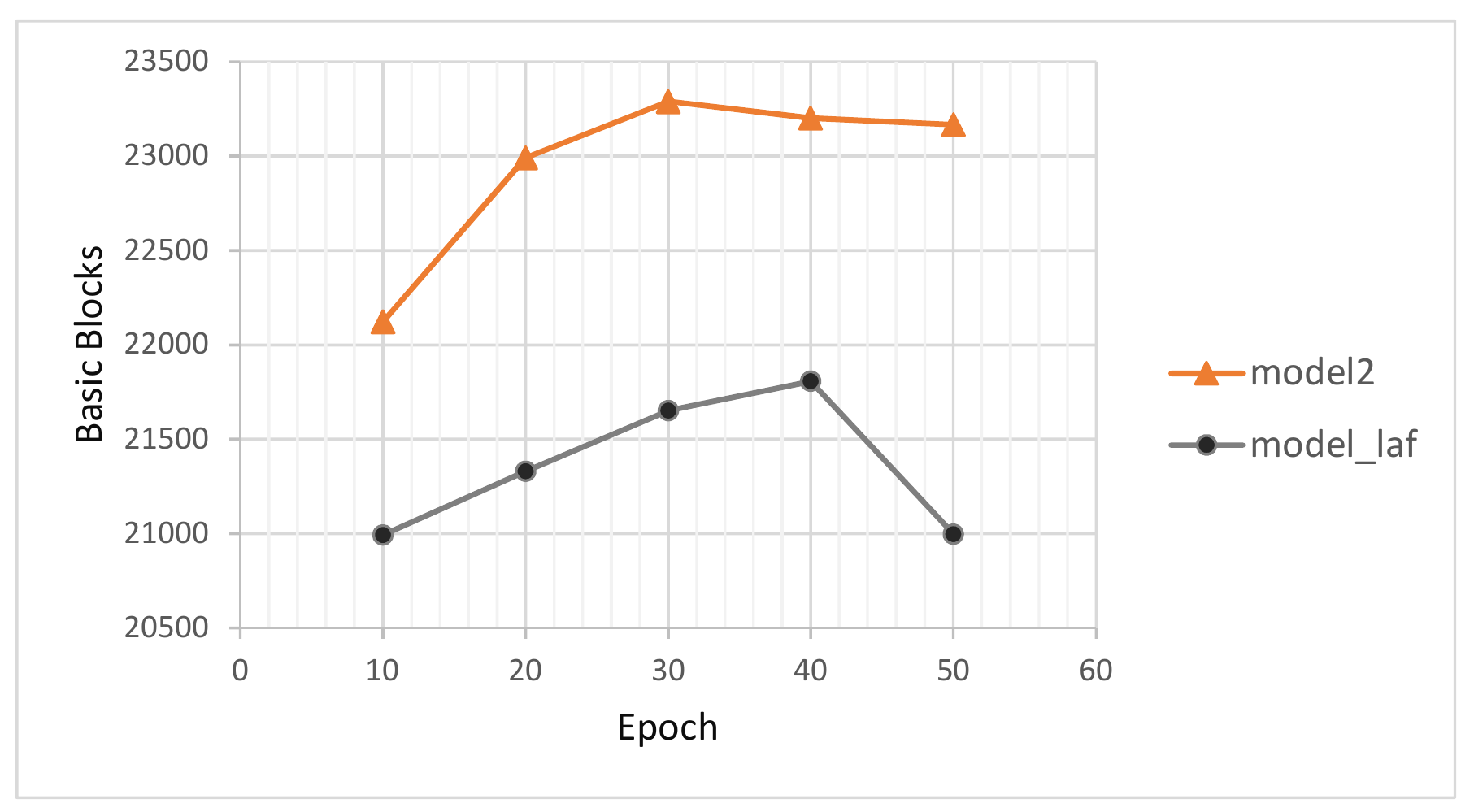}
	\caption{Code coverage per epoch for model 2 and model \textit{laf} during the training process.}
	\label{fig:cmp-laf-model2}
\end{figure}

Another parameter that affects code coverage is the number of training epochs. We compare code coverage of model 2 and Learn\&Fuzz \cite{Godefroid:2017:LML:3155562.3155573} in five different epochs include epochs 10, 20, 30, 40 and 50. To do this, the model checkpoint at the specific epoch is used to generate 1,000 PDF files and then code coverage is obtained, again by running the test suits on MuPDF viewer \cite{MuPDF2018}. The result is shown in Figure \ref{fig:cmp-laf-model2}. In both models, the coverage increases in early epochs of training and then begins to decrease. This occurs because the training improves the model in the early epochs then the model behavior shows a bias on trainset, and it may be overfitting. However, in general, it seems that there is no definite relationship between the number of training epochs and code coverage. The remarkable point is that model 2 has a higher code coverage in all training epochs.

\subsection{Neural Fuzz Testing}
\noindent In the fourth and the last experiment, we put MuPDF \cite{MuPDF2018} on the real fuzz testing. We generate 10,000 PDF files using each of \textit{DataNeuralFuzz} and \textit{MetadataNeuralFuzz} algorithms and then use \iust{} to do fuzz testing. Besides fuzzing with our neural fuzzing algorithms, we do fuzz testing by FileFuzz \cite{Sutton:2007:FBF:1324770}, a simple mutation-based file format fuzzer, and Learn\&Fuzz (i.e. \textit{SampleFuzz} algorithm) \cite{Godefroid:2017:LML:3155562.3155573}. In all experiments, we use \textit{host1\_max} as the host file or initial seed in the case of FileFuzz. We fuzzed MuPDF viewer with total of 40,000 ($40K$) PDF files in this experiment. 

Table \ref{table:neuralfuzz-running-config} shows the input and constant values which are set for \textit{DataNeuralFuzz} and \textit{MetadataNeuralFuzz} algorithms to generate test data among the available values. Table \ref{table:fuzzing-result} shows the code coverage results of various fuzz testing methods include Learn\&Fuzz \cite{Godefroid:2017:LML:3155562.3155573} and FileFuzz\cite{Sutton:2007:FBF:1324770}. Finally, table \ref{table:neuralfuzz-improvements} shows the difference between code coverage of our proposed method and four other known file format fuzzers: Learn\&Fuzz, AFL \cite{Zalewsky2013}, Augmented-AFL \cite{DBLP:journals/corr/abs-1711-04596}, and FileFuzz. Augmented-AFL has been introduced recently by Microsoft research as an improvement for AFL. The following results are observed.

\begin{table}[]
	\centering
	\caption{Input and constant values for our neural fuzz algorithms during test data generation.}
	\resizebox{0.85\textwidth}{!}{%
		\begin{tabular}{@{}lll@{}}
			\toprule
			Input / constant             & Available values    & Selected values                     \\ \midrule
			Learned model M               & 1, 2, 3, 4, laf     & 2                                   \\
			Sequence prefix P             & String constant     & Randomly selected from the test set \\
			Diversity D                   & (0, $+\infty$)      & 1                                   \\
			Fuzzing rate FR               & (0, 1{]}            & 0.1                                 \\
			End token ET                  & String constant     & endobj                              \\
			Binary token BT               & String constant     & stream                              \\
			(a, b)                        & (len(p), $+\infty$) & (450, 550)                          \\
			$\alpha$ in DataNeuralFuzz  & (0, 1)              & 0.5                                 \\
			$\beta$ in MetadataNeuralFuzz & (0, 1)              & 0.9                                 \\ \bottomrule
		\end{tabular}%
	}

	\label{table:neuralfuzz-running-config}
\end{table}

\begin{table}[]
	\centering
	\caption{Results of fuzz testing with 10,000 PDF files for each algorithm.}
	\label{table:fuzzing-result}
	\resizebox{0.85\textwidth}{!}{%
		\begin{tabular}{@{}lllll@{}}
			\toprule
			Algorithm & Basic block coverage & Percent & Line coverage & Percent \\ \midrule
			DataNeuralFuzz & 23,719 & 19.36 & 18,673 & 20.81 \\
			MetadataNeuralFuzz & 22,583 & 18.43 & 17,894 & 19.95 \\
			SampleFuzz \cite{Godefroid:2017:LML:3155562.3155573} & 20,957 & 17.10 & 16,793 & 18.72 \\
			RandomFuzz (FileFuzz \cite{Sutton:2007:FBF:1324770}) & 7,563 & 6.17 & 5,002 & 5.58 \\ \bottomrule
		\end{tabular}%
	}
\end{table}

\begin{table}[]
	\centering
	\caption{Improvement of proposed algorithms code coverage in comparison with existing fuzzers. Each number shows the difference between code coverage of algorithms in its column with its row. All value is in percent.}
	\label{table:neuralfuzz-improvements}
	\resizebox{0.85\textwidth}{!}{%
		\begin{tabular}{@{}lll@{}}
			\toprule
			Algorithm / fuzzer    & DataNeuralFuzz & MetadataNeuralFuzz \\ \midrule
			SampleFuzz \cite{Godefroid:2017:LML:3155562.3155573}           & +2.26          & +1.33              \\
			AFL \cite{DBLP:journals/corr/abs-1711-04596}                  & +7.73          & +6.80              \\
			Augmented-AFL \cite{DBLP:journals/corr/abs-1711-04596}         & +7.56          & +6.63              \\
			RandomFuzz (FileFuzz \cite{Sutton:2007:FBF:1324770}) & +13.19         & +12.26             \\ \bottomrule
		\end{tabular}%
	}
\end{table}

\begin{itemize}
	\item{\textit{MetadataNeuralFuzz} code coverage is less than \textit{DataNeuralFuzz}. As we have said already, manipulating a small part of the file format may make it wholly invalid, and hence the file is rejected by the parser as soon as and lead to low code coverage. However, changing the data within file affects the rendering stage of the file execution. The results prove that both algorithms act as we expected. The one fuzzes format, and the another fuzzes data.
	}
	
	\item{Both \textit{DataNeuralFuzz} and \textit{MetadataNeuralFuzz} have covered more basic blocks (of course more lines) of MuPDF viewer code than \textit{SampleFuzz} \cite{Godefroid:2017:LML:3155562.3155573}. That shows NLMs with RNNs outperforms encoder-decoder models in fuzz testing. Another interpretation is that hybrid test data generation beats the generation-based methods.
	}
	
	\item{Our hybrids test data generation methods also outperform mutation based fuzzers such as AFL and AugmentAFL, as shown in table \ref{table:neuralfuzz-improvements}. The code coverage for AFL and Augmented-AFL have taken from \cite{DBLP:journals/corr/abs-1711-04596} as benchmarks. }
	
	\item{The advantage of intelligence algorithms versus random mutation based on the test data generation part of fuzz testing is obvious. The random algorithms cannot access the high code coverage in the complex input structures. The coverage of the \textit{DataNeuralFuzz} algorithm is more than three times the algorithm using in FileFuzz \cite{Sutton:2007:FBF:1324770}.}
	
	\item{Although we have improved the code coverage of MuPDF viewer \cite{MuPDF2018} during fuzz testing somewhat, as we see in table \ref{table:fuzzing-result}, the percentage of covered code is still below 25\%. This means the most of the viewer codes are not executed, and it is not good news. On the other hand, we should know that MuPDF viewer can parse and play different file format such as XPS. This means that part of the not executed code is used when inputs are in such formats. Thus, we do not expect to run them just by generating and injecting PDF files.
	}
\end{itemize}

\subsection{Faults and Vulnerabilities}
\noindent The best metric which can be used to evaluate a fuzzer is the number of faults and vulnerabilities found during fuzz testing. We did not see any errors in reports generated by Application Verifier \cite{ApplicationVerifier} after each test execution. Given that we tested the final version of the MuPDF software \cite{MuPDF2018}, it is assumed that most of its errors are fixed in the trial versions, and thus it will be difficult to find the new fault. On the other hand, MuPDF is software under active development, and it has great developers and user community that makes it robust software. However, the \textit{DataNeuralFuzz} algorithm detected several uses of unsafe functions and reported them as a security warning. 

It seems that Application Verifier \cite{ApplicationVerifier} when running on Windows 10 x64 unable to detect 32bit applications memory errors. We try to fuzz testing a trivial 32bit application with known fault, but ApplicationVerifier does not report anything. The 64bit application but do not have such problems, and their faults are detected by ApplicationVerifier. Hence we tested both 32bit and 64bit version of MuPDF viewer \cite{MuPDF2018}. IUST DeepFuzz opens SUT with a test data and closes it after a fixed time, the try to inject next test data in test suit. At our configuration each test suit contains 10,000 test data take about 28 hours to be processed. Fuzzing is kind of stress testing typically done in several days or weeks to find faults and vulnerabilities. We are planning to test MuPDF on more massive test suits in order of 100,000 ($100K$) PDF files and more which probably can break MuPDF.

\section{Related Works}
\noindent In this section, we discuss some related works in fuzzing and explain their existing problems concerning test data generation. According to the test data generation methods, fuzzers are categorized as Mutation-based and generation-based \cite{Mcnally2012,CHEN2018118, Li2018}. Various techniques are applied to both methods to improve them. Most of these techniques have focused on artificial intelligence algorithms. 

\begin{enumerate}[I.]
	\item{\textbf{Mutation-based Fuzzing.} In mutation-based, one or more valid input data is used as the initial seed. This seed then mutates to produce another test data. It is easy to construct mutation-based fuzzer and generate mal-formed test data with it. In this case, there is no necessity for a prior understanding of the input data structure. The drawback of the mutation-based method is that this method depends on the variation of the initial seed. Without different sample inputs, mutation-based fuzzers does not achieve high code coverage \cite{Miller2007} which shows the importance of the initial seed in the mutation-based methods. AFL \cite{Zalewsky2013} and FileFuzz \cite{Sutton:2007:FBF:1324770} are examples of the mutation-based fuzzers. }
	
	\item{\textbf{Generation-based Fuzzing.} The generation-based method generates test data entirely random or from a formal description such as grammar, template, or model. The latest uses the input format specifications to construct a generative model. This method is most often applied to the formats that some documentation available for them. Usually, it achieves a higher code coverage, in comparison with mutation-based fuzzers \cite{Miller2007}. However, as we said, a lot of time and money should be spent to get the specifications of the file format fully understood and build a proper grammar, template, or model for it. SAGE \cite{SAGE2012} and Peach \cite{Peach} are examples of generation-based fuzzer. There are also hybrid methods which utilize the features of both approaches. IUST DeepFuzz proposed in this paper is a hybrid fuzzer that generate structured textual data by a generative model and unstructured binary data by mutations.}
	
	\item{\textbf{Evolutionary Fuzzing.} First attempts to bring intelligence to fuzzing were done by applying evolutionary algorithms such as genetic \cite{DeMott2007}. An evolutionary fuzzer receives feedback from runtime information, typically code coverage information, and adds those test data that leads to the new execution paths into a queue. After that, when the fuzzer wants to generate test data, it only mutates the test data which exist in the queue, hoping to be able to run new parts of the code. AFL \cite{Zalewsky2013} is the state of the art evolutionary file format fuzzer works exactly like those above. By using the feedback taken from previous runs, AFL can choose better test data; however, it mutates them randomly. As a result, a large number of duplicate test data will be generated that do not necessarily affect the testing criteria including code coverage. On the other hand, in complex input structures, changing some critical parts causes the input test data to be rejected by the parser at the initial stage of parsing. So, we need a mechanism to inform fuzzer where to mutate (which offset) the input file.}
	
	\item{\textbf{Deep Learning in Mutation-based and Evolutionary Methods.}
		Augmented-AFL \cite{DBLP:journals/corr/abs-1711-04596}, as an improvement patch for AFL \cite{Zalewsky2013},  tries to find suitable places for mutating bytes using deep learning techniques. After Augmented-AFL created a new test data, it queries a model to see whether the generated test data is good enough or not? This method increases the test speed, but a large amount of data are rejected by the model (veto) while being produced. Also, Augmented-AFL does not exhibit significant improvements in code coverage for MuPDF parser \cite{MuPDF2018}. It seems for the applications with complex input structure mutation-based methods cannot rich high code coverage.
	}
	
	\item{\textbf{Deep Learning in Generation-based Methods.}
		Applying the neural-network-based statistical learning to automatically generate input grammars from sample inputs was initially proposed by Godefroid et al. \cite{Godefroid:2017:LML:3155562.3155573}. They also presented an algorithm for generating fuzzing inputs. The main idea of the work is to learn a generative model over a set of PDF files \cite{Incorporated2006}. To this aim, they used a kind of sequence to sequence architecture \cite{NIPS2014_5346,DBLP:journals/corr/ChoMGBSB14} which is originally used in mapping two sequences from different domain together, e.g., the task of machine translation. They called their method Learn\&Fuzz. Throughout the paper, we argued some weaknesses of the Learn\&Fuzz method and provided some solutions for them. Based on this work, Cummins et al. introduced DeepSmith \cite{Cummins:2018:CFT:3213846.3213848} that uses LSTM architecture of RNN \cite{doi:10.1162/neco.1997.9.8.1735} to model the program code. They applied the tool for fuzzing compiler of OpenCL programming language. Their model is not a hybrid model and can use only to generate textual test data.}
	
\end{enumerate}

\section{Conclusion}
\noindent This article is aimed at the introduction of a new intelligent test data generation technique for complex input structures such as PDF files. Deep neural language models, built by recurrent neural networks, could be best applied to learn the structure of complex input files as a sequence of symbols. Textual sections of input files could be simply learned. However, it is a difficult job to learn the format of a binary section. To resolve the difficulty, we suggest to temporarily delete binary sections and substitute these sections with a specific token. After the training phase is completed and when the learned model is applied to generate test data, the tokens are replaced with the mutated form of the deleted sections.  To improve the fuzzer efficiency, we fuzz both the data and meta-data when applying the learned model to generate new input files as test data. We believe that both of the presented algorithms are required when fuzz testing is done regardless of the code coverage. Neural fuzzing algorithms are designed to test different parts of the program. \textit{MetadataNeuralFuzz} tests parser of file format and \textit{DataNeuralFuzz} tests renderer of file format.

Test data generator is the most important module in fuzzers. Providing an automatic test data generator that can achieve high code coverage in the software under test, especially the targets with complex input structure, is essential to find faults. Generation-based and mutation-based methods have been successfully applied to generate test data for fuzzing. However, the former is not fully automatic, and the latter suffers from poor code coverage. 

To address these problems, we propose an approach based on NLMs and deep learning techniques. Our hybrid test data generation method automatically learns the structure of the input file and then generates new diverse test data by fuzzing both textual and binary parts of the input format. As the method intelligently determines the location of fuzz and the value which should be used to fuzz, it can be promisingly applied for testing complex targets. 

We conducted our experiments on a complex file format, i.e., PDF, and the results confirm the significant improvement of the code coverage and the accuracy of our proposed approach compared to the previous methods. Besides the general conclusion, our analysis reveals some worthwhile empirical facts, most notably:

\begin{itemize}
	\item{Hybrid test data generation for fuzzing both textual and binary parts of complex input structures, increase code coverage of the SUT.}
	
	\item{It is widely recognized that bidirectional LSTM as an LM can obtain more accuracy and less error on the same dataset. However, it is observed that simpler NLMs such as unidirectional LSTM without dropout, e.g., model 2 in this paper, can outperform more complex method in the code coverage. A similar result is reported in \cite{DBLP:journals/corr/abs-1711-04596}.}
	
	\item{The incremental update process based on a PDF file with high code coverage results in more code coverage.}
	
	\item{Despite providing relatively higher code coverage than random and existing intelligence fuzzers, our proposed fuzzer can be improved to provide higher coverage for complex input structures such as PDF file structure.}
	
\end{itemize}

There is a vast area of future work on this topic. One is to use other powerful deep learning models such as generative adversarial networks (GANs) \cite{NIPS2014_5423} to generate the test data. Another direction is to apply these models for generating test data in other types of fuzzers such as network protocol fuzzers. In order to produce more effective test data, we intend to add a feedback loop to IUST DeepFuzz aimed at receiving the runtime information and fine tuning the learned model. There are parts of code in SUT that handle user interactions. Fuzzers, such as AFL \cite{Zalewsky2013} and IUST DeepFuzz, however, do not utilize user interaction parts for fuzzing and does not support the execution of these parts of the code. For the time being, we are planning to support the automation of user interactions with SUT.

\section*{References}
\bibliography{neuralfuzzing}
\end{document}